\documentclass[fleqn,usenatbib]{mnras}

\usepackage[T1]{fontenc}

\DeclareRobustCommand{\VAN}[3]{#2}
\let\VANthebibliography\thebibliography
\def\thebibliography{\DeclareRobustCommand{\VAN}[3]{##3}\VANthebibliography}

\usepackage{graphicx}	%
\usepackage{amsmath}	%
\usepackage{gensymb}
\usepackage{multirow}
\usepackage{bm}
\usepackage{caption}
\usepackage{subcaption}
\usepackage{array}
\usepackage{newtxtext,newtxmath, amsmath}

\newcommand{\matr}[1]{\bm{#1}}

\title[offset between Sgr dSph and M54]{Offset of M54 from the Sagittarius Dwarf Spheroidal Galaxy}

\author[Z. An et al.]{
    Zhaozhou An,$^{1}$\thanks{E-mail: za@andrew.cmu.edu}
    Matthew G. Walker,$^{1}$
    Andrew B. Pace$^{1}$
    \\
    $^{1}$McWilliams Center for Cosmology, Carnegie Mellon University, 5000 Forbes Ave, 15213, USA\\
}

\date{Accepted XXX. Received YYY; in original form ZZZ}

\pubyear{2024}

\begin{document}
\label{firstpage}
\pagerange{\pageref{firstpage}--\pageref{lastpage}}
\maketitle

\begin{abstract}
    We present results from simultaneous modeling of 2D (projected along the line of sight) position, proper motion and line-of-sight velocity for \textit{Gaia}- and APOGEE-observed stars near the centre of the Sagittarius (Sgr) dwarf spheroidal galaxy.  We use a mixture model that allows for independent sub-populations contributed by the Sgr galaxy, its nuclear star cluster M54, and the Milky Way foreground.
    We find an offset of $0.295\pm 0.029$ degrees between the inferred centroids of Sgr and M54, corresponding to a (projected) physical separation of $0.135\pm 0.013$ kpc.  The detected offset might plausibly be driven by unmodelled asymmetry in Sgr's stellar configuration; however, standard criteria for model selection favour our symmetric model over an alternative that allows for bilateral asymmetry.  We infer an offset between the proper motion centres of Sgr and M54 of $[\Delta\mu_{\alpha}\cos\delta,\Delta\mu_{\delta}]=[4.9, -19.7] \pm [6.8, 6.2]$ $\mu$as yr$^{-1}$ ($[0.61, -2.46] \pm [0.85, 0.77] $ km s$^{-1}$), with magnitude similar to the covariance expected due to spatially-correlated systematic error.  We infer an offset of $4.1\pm 1.2$ km s$^{-1}$ in line-of-sight velocity.  
    Using inferred values for the systemic positions and motions of Sgr and M54 as initial conditions, we calculate the recent orbital history of a simplified Sgr/M54 system, which we demonstrate to be sensitive to any line-of-sight distance offset between M54 and Sgr, and to the distribution of dark matter within Sgr.  
\end{abstract}


\section{Introduction}
\label{sec:intro}

Observations reveal that a significant fraction of galaxies host dense nuclear star clusters (NSCs) at their centres \citep{cote_acs_2006, turner_acs_2012,2014ApJ...791..133B,den_brok_hstacs_2014,2019ApJ...878...18S,2020A&ARv..28....4N}.  However, the
formation of NSCs is still a debated question.  Possible explanations range from in situ star formation at the galactic centre
\citep{2006ApJ...642L.133B,2007PASA...24...77B, 2015ApJ...812...72A} to the infall of globular clusters (GCs), whose orbits decay and spiral to the centre due to
dynamical friction \citet{1975ApJ...196..407T, 1993ApJ...415..616C, 2014MNRAS.444.3738A,2014ApJ...785...71G}.

Some NSCs are found to host metal-poor stars \citep{1985AJ.....90.1927R,
    2010A&A...520A..95C, 2015ApJ...809..143D}, which can be naturally explained by
old, metal-poor GCs bringing those stars into the nucleus.  Deficits of bright non-nuclear GCs near the central regions of elliptical galaxies also seem to indicate
that the central nuclei formed via the orbital decay of massive clusters
\citep{capuzzo-dolcetta_globular_2009, lotz_dynamical_2001}. In early-type galaxies with stellar mass $\lesssim 10^9\, \mathrm{M_{\odot}}$, the NSCs and GC systems have similar occupation distributions and comparable total masses \citep{2019ApJ...878...18S}, consistent with NSC formation by infalling GCs.  

However, many NSCs also contain young stellar
populations \citep{seth_clues_2006, walcher_stellar_2006, nguyen_extended_2014,
    bender_hst_2005, paumard_two_2006}, suggestive of in situ star formation within the host galaxy.  Thus, it seems that both the
GC infall and in situ star formation contribute to NSC formation.  Indeed some NSCs display both young and old stellar populations with different morphology
and kinematics, which indicates  different origins for different
components and multiple star formation episodes \citep{seth_clues_2006,
    2006AJ....132.1074R, walcher_stellar_2006}. For example, \citet{2011MNRAS.418.2697H} show
that it is necessary to consider both stellar dynamical mergers and in situ
star formation to reproduce observations of the NSCs at the
centres of NGC 422 and M33. \citet{2015ApJ...812...72A} use a semi-analytical
galaxy formation model that includes both dynamical-friction-driven migration
of stellar clusters and star formation triggered by infalling gas to simulate
the co-evolution of NSCs and central black holes, finding that in situ star formation contributes a
significant fraction (up to $\sim 80\%$) of the total mass of NSCs.

Considering the various pathways of NSC formation, it is useful to make 
 detailed comparisons between the spatial and chemodynamical properties of NSCs and their host galaxies.  Notably, some works find evidence for offsets between the centres of some NSCs and the centres of their hosts.   
For example, \citet{binggeli_off_center_2000} search for off-centre nuclei in 78
nucleated dwarf elliptical galaxies of the Virgo Cluster, finding that roughly
$20\%$ of the sample are significantly off-centre.
\citet{cote_acs_2006} study $100$
early-type galaxies of the Virgo Cluster, identifying five candidates with nuclei offset from the host galaxy photocentre; they note, however, it is possible that some of 
these cases are due to non-nuclear globular clusters that happen to be projected close to the galaxy photocentre.  Such offsets, if real, may indicate oscillations of the NSC that may help
constrain, e.g., the mass of a central black hole \citep{1998MNRAS.299..111T} and/or other physical processes like counterstreaming instability \citep{2004ApJ...603L..25D}.

Messier 54 (M54 also known as NGC 6715), the NSC within the Sagittarius (Sgr) dwarf spheroidal (dSph) galaxy, offers a unique opportunity to observe detailed spatial and chemodynamic properties of individual stars within a NSC and its local environment.  Originally discovered in a line-of-sight velocity survey of the Galactic bulge \citep{1994Natur.370..194I}, Sgr lies at a distance of $\sim 26$ kpc from the Sun,  $\sim 18$ kpc from the Galactic centre, and is the progenitor of the prominent Sagittarius stream, stellar debris from the ongoing process of Sgr's tidal disruption within the Milky Way \citep{2001ApJ...547L.133I,2003ApJ...599.1082M,2006ApJ...642L.137B}.  The central region of Sgr, which includes M54, displays a complex formation history, with various photometric and spectroscopic studies identifying old ($\gtrsim 10$ Gyr), intermediate-age ($\sim 4-6$ Gyr) and young ($\lesssim 3$ Gyr) stellar populations with a corresponding range of metallicity from -1.8$\lesssim$ [Fe/H]$\lesssim +0.6$ \citep[e.g.][]{1995AJ....109.1086S,2007ApJ...667L..57S,2011ApJ...743...20S,2019ApJ...886...57A}.  

This complexity extends to the observed stellar kinematics.  The old, metal-poor stars have a line-of-sight velocity dispersion profile that declines steeply with cluster-centric radius \citep{2009ApJ...699L.169I}, then gradually rises near M54's nominal tidal radius \citep{2008AJ....136.1147B}.  Over the same region, the relatively metal-rich stars display approximately constant velocity dispersion, as is characteristic of the Milky Way's dwarf spheroidal satellite galaxies \citep{2008AJ....136.1147B}.  Even outside the nucleus, the main body of Sgr itself exhibits complex internal stellar kinematics, with a central bar-like structure that connects to `tails' of escaping stars, a bound, rotating inner core,  and apparent expansion along the long axis that characterises an overall triaxial morphology   
\citep{2021ApJ...908..244D}.  

Given this complexity, it is difficult to disentangle M54 from its Sgr host.  Analysing the spatial distribution of a large photometrically-selected sample of M giant stars from the Two Micron All Sky Survey,  \citet{2003ApJ...599.1082M} find an overdense `cusp' at the position of M54; however, they argue that this feature belongs to Sgr and not to M54, whose metal-poor stars are too blue to be included in the M giant sample.  Analysing metal-rich and metal-poor populations selected from optical photometry, \citet{monaco_central_2005} conclude similarly, finding that the metal-rich stars (presumed to belong to Sgr) exhibit an overdense cusp at the same location as the cusp seen in metal-poor stars (presumed to belong to M54).  However, a MUSE spectroscopic survey of M54 by \citet{2019ApJ...886...57A} finds that the metal-rich stars can be separated into young ($\sim 2$ Gyr) and intermediate-age ($\sim 4$ Gyr) populations, associating the more compact, younger subpopulation with late star formation within M54, and the more extended, intermediate-age sub-population with the Sgr host.

If the old, metal-poor component of M54 is an example of a GC that fell to the centre of its host galaxy, then its properties may also help to constrain the spatial distribution of dark matter within Sgr.  Various dynamical and N-body studies demonstrate that the efficiency and outcome of dynamical friction depend on the spatial distribution of dark matter within the host galaxy \citep[e.g.,][]{2006MNRAS.368.1073G,2006MNRAS.373.1451R,2010ApJ...725.1707G,2020MNRAS.491.3336M}.  If the host dark matter halo has the central `cusp' that characterises cold dark matter halos \citep[][`NFW' hereafter]{1997ApJ...490..493N}, dynamical friction operates efficiently to drag a massive GC to the centre in a fraction of a Hubble time; if instead the host halo has a central `core' of uniform density, the infalling cluster tends to stall near the core radius, where the halo becomes effectively `bouyant' \citep{2012MNRAS.426..601C,2022ApJ...926..215B}.  Using N-body models to simulate specifically the infall of M54 within Sgr, \citet{2008AJ....136.1147B} find that for a variety of initial conditions, if the Sgr host has an NFW-like cusp, then M54 sinks efficiently (in $\lesssim 3$ Gyr) to the very centre of Sgr, where it remains virtually at rest, with a systemic velocity offset of $\lesssim 1-2$ km s$^{-1}$.  These results are broadly consistent with those of \citet{2023MNRAS.523.2721H}, who also find that a more (than NFW) steeply cusped dark matter halo in Sgr would result in a NSC that (internally) rotates faster and is morphologically flatter than M54.

Thus the observed properties of the Sgr/M54 system impact our understanding not only of NSC formation, but also the nature of dark matter.  In any case, the degree to which M54 and the centre of Sgr are either distinct or coincident in phase space remains an interesting open question.  The answer is complicated for several reasons.  First, as mentioned above, the central regions of Sgr display a mixture of several stellar populations with different ages, metallicities, structural parameters and kinematics, and the assignment of a given population to one or the other object can be ambiguous.  Second, published measurements of centroids, distances and systemic velocities often pertain to just one of Sgr or M54.  For example,  the infrared M-giant sample of \citet{2003ApJ...599.1082M} is relatively insensitive to the metal-poor population in M54; conversely, Hubble Space Telescope and the MUSE observations of M54 typically lack the sky coverage to map Sgr 
 over a significant fraction of its core region  \citep[e.g.,][]{2011ApJ...743...20S,2019ApJ...886...57A}.  As a result, comparisons of published measurements for Sgr and M54 must contend with systematic errors (e.g., zero-point offsets) introduced by the use of different data sets.  

A notable exception is the recent study by \citet{2021ApJ...908..244D}, which uses the Gaia (DR2) catalogue, supplemented with training data from spectroscopic (line-of-sight velocity) catalogues, to construct a catalog of 6D phase-space coordinates for a final sample of $\sim 1.2\times 10^5$ stars within $\sim 6^{\circ}$ of the Sgr centre.  With this catalogue, they iteratively measure the centre-of-mass coordinates of a Sgr sample that excludes all stars within the fiducial tidal radius of M54, and separately the center-of-mass coordinates for the M54 sample that includes only the stars excluded from the Sgr sample. 
 They find a statistically significant offset only among the plane-of-sky centroids, which they discount as likely driven by systematic error due to  incompleteness that  results from the Gaia scan pattern that is apparent in their sample (their Figure 4).  

Here we take an approach that is complementary to that of \citet{2021ApJ...908..244D}.  Rather than adopting hard cuts in colour/magnitude or sky position to separate M54 and Sgr samples, we use the mixture models to analyse the 5D distribution of sky position, proper motion and line-of-sight velocity, the phase-space coordinates for which homogeneous measurements from Gaia EDR3 \citep{2021A&A...649A...1G} and/or the Apache Point Observatory Galactic Evolution Experiment \citep[APOGEE;][]{2017AJ....154...94M} are available for large numbers of stars in the Sgr.  The mixture modelling lets us obtain simultaneous estimates for the 5D centres of the Sgr and M54 populations, providing a direct inference about any offset in these dimensions of phase space.  Any detected offset can then be used to inform subsequent models for the formation and evolution of M54 as the NSC within Sgr.  Furthermore, since our mixture models do not operate separately on pre-selected M54 and Sgr samples, the result can be used to infer, rather than assume, the colour/magnitude distributions traced separately by the two objects.  

We describe the data selection procedure in Section~\ref{sec:data}. In Section~\ref{sec:method}, we describe the construction of the mixture model.  In Section \ref{sec:results}, we present results from our modelling. In Section~\ref{sec:discussion}, we compare our results to previous work and discuss the robustness of our results, the colour/magnitude distribution of the M54, and the recent orbital history of the Sgr/M54 system.  In Section~\ref{sec:conclusions}, we summarise our conclusions.

\section{Data}
\label{sec:data}

This work is based on data presented in the Early Data Release 3 \citep[EDR3,][]{2021A&A...649A...1G} of the Gaia mission \citep{2016A&A...595A...1G}, and on line-of-sight velocity data presented by APOGEE Data Release 17 \citep[DR17;][]{2022ApJS..259...35A}.  %
\textit{Gaia} EDR3 consists of astrometry and photometry for $1.8$ billion sources brighter than magnitude $G\approx 21$.  Compared with \textit{Gaia} DR2, \textit{Gaia} EDR3 improves the parallax precision by $30\%$ and increases proper motion precision by a factor of 2. At magnitude $G=17$, it provides typical uncertainties of $0.07\,\mathrm{mas\ yr^{-1}}$ in proper motion, $0.07\, \mathrm{mas}$ in parallax angle, $0.05\, \mathrm{mas}$ in position, $1\, \mathrm{mmag}$ in G-band magnitude, $12\, \mathrm{mmag}$ in $\mathrm{G_{BP}}$ magnitude, and $6\, \mathrm{mmag}$ in $\mathrm{G_{RP}}$ magnitude \citep{2021A&A...649A...1G}. %
APOGEE DR17 is the continuation of the Apache Point Observatory Galactic Evolution Experiment \citep[APOGEE,][]{2017AJ....154...94M} and Data Release 17 of APOGEE-2/SDSS-IV  \citep{2016AN....337..863M}.  APOGEE spectroscopic catalogues include stellar-atmospheric parameters and line-of-sight velocities measured from twin, multi-plexed, near-infrared, high-resolution spectra covering both the northern and southern sky \citep{2022ApJS..259...35A, 2019PASP..131e5001W}. 
 The extracted line-of-sight velocities have a median reported uncertainty of $0.015 \, \mathrm{km\ s^{-1}}$.

\subsection{\textit{Gaia} EDR3 Data Selection}
\label{sec:gaia_data_selection}

\begin{figure}
    \centering
    \includegraphics[width=\columnwidth]{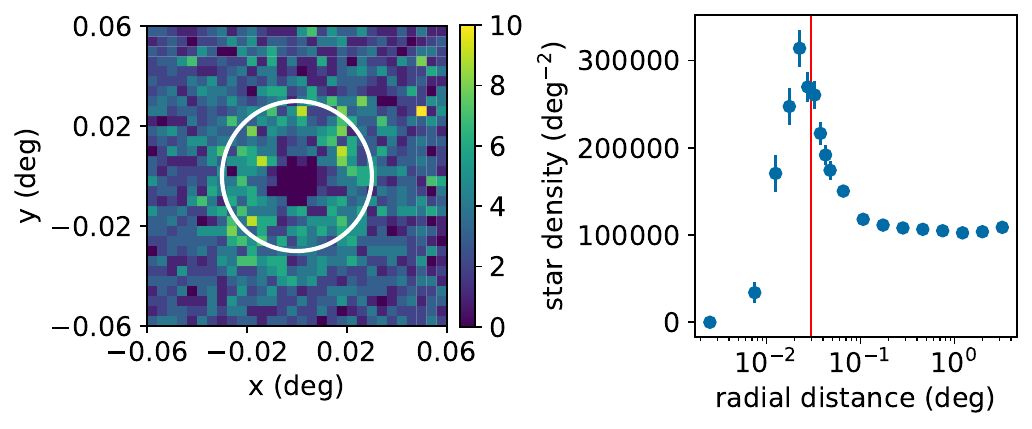}
    \caption{Projected density of stars around the centre of M54, from \textit{Gaia} EDR3, with filters $\texttt{phot\_g\_mean\_mag}<20$, $\texttt{astrometric\_excess\_noise\_sig}<2$ and $\texttt{phot\_bp\_mean\_mag}$ and $\texttt{phot\_rp\_mean\_mag}$ exist.  The left panel shows the 2D density field and the color is coded by the number of the stars in the bin; the right panel shows the mean density within circular annuli.  The white circle and the red line denote a radius of $0.03^{\circ}$, inside which we mask the sample.}
    \label{fig:spatial2d_center}
\end{figure}

We begin by considering all sources from Gaia EDR3 that lie within $4^{\circ}$ of a fiducial center of M54 \citep[$\alpha_{2000}=283.76353^{\circ},\delta_{2000}=-30.477006^{\circ}$;][]{2019ApJ...886...57A}.  Within this region, one immediately obvious systematic effect is the tendency for sources near the centre of M54 to have relatively poor astrometric solutions due to crowding.  Figure \ref{fig:spatial2d_center} demonstrates that application of the standard filter for selecting astrometrically well-behaved sources, ${\verb|astrometric_excess_noise_sig|}<2$ \citep{2012A&A...538A..78L} removes almost all sources within $\sim 1$ arcmin, or $\sim 2$ half-mass radii of M54 ($r_h=0.43\pm 0.08$ arcmin; \citealt{2019ApJ...886...57A}).  To avoid this 
selection bias, we mask the entire region within $\sim 0.03^{\circ}$ ($1.8$ arcmin) of the fiducial centre of M54. We note that special Service-Interface Function (SIF) images of dense area, which includes the centre of the Sgr dSph, are acquired in the sky-mapper CCDs \citep{2016A&A...595A...1G}, and that can be used to analyse the centre of the Sgr dSph. One example is the data for $\omega$ Centauri in \citet{2023A&A...680A..35G}. However, these special SIF data are not part of Gaia DR3.

\begin{figure}
    \centering
    \includegraphics[width=\columnwidth]{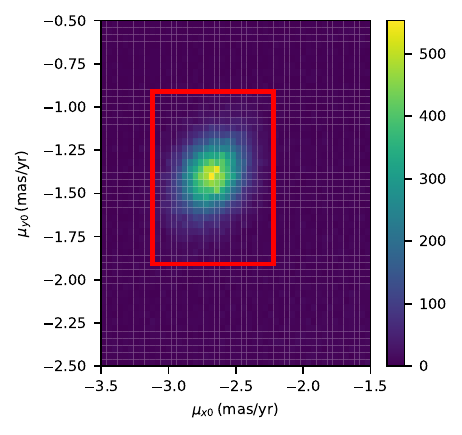}
    \caption{Distribution of proper motions of stars within $4^{\circ}$ of the nominal center of M54, from \textit{Gaia} EDR3, with all filters (except the proper motion filter) listed in  Sec~\ref{sec:gaia_data_selection} applied.  Color is coded by the number of stars within each pixel. The red rectangle indicates our proper motion filter.}
    \label{fig:pm_selection}
\end{figure}

Next, following \citet{2022MNRAS.tmp.2672V}, we impose a magnitude cut at $G<17.3$ in order to ensure high-quality astrometry and homogeneous coverage over the region of interest.  Then we cut on proper motion based on a filter which is chosen by eye to remove as many MW stars as possible, without losing stars from M54 and Sgr, as shown in Fig~\ref{fig:pm_selection}.
We make additional cuts on parallax to remove stars with distance less than $20$ kpc and apply data quality flags.  Our complete initial selection uses the following criteria:

\smallskip

$ 0.03 \degree \leq \sqrt{x^2+y^2} \leq 4 \degree$

${\verb|phot_g_mean_mag|}<17.3$,

${\verb|astrometric_excess_noise_sig|}<2$,

${\tt parallax}<0.05+3 \times {\tt parallax\_error}$,

${\tt pmra}\in [-3.12, -2.22]$, ${\tt pmdec} \in [-1.91, -0.91]$

${\tt ipd\_frac\_multi\_peak} < 2$

${\tt ruwe} < 1.3$

${\tt duplicated\_source} = {\tt False}$

$|{\tt C^\star}|<3{\tt \sigma_{C^\star}}$.

\smallskip

\noindent Here, $(x,y)$ are rectilinear coordinates obtained via gnomonic projection of the Gaia-measured sky coordinates, assuming an origin at the previously-published center of M54 \citep[$\alpha_{2000}=283.76353^{\circ},\delta_{2000}=-30.477006^{\circ}$;][]{2019ApJ...886...57A}.  ${\tt C^\star}$ is the corrected BP and RP flux excess factor, defined in Equation (6) of \cite{2021A&A...649A...3R} and ${\tt \sigma_{C^\star}}$ is the
$1\sigma$ scatter, calculated according to Equation (18) of
\citet{2021A&A...649A...3R}.  Using the `astrometric fidelity' parameter of \citep{2022MNRAS.510.2597R}, all of the sources in our final sample have $\tt{fidelity}>0.5$ \citep{2022MNRAS.510.2597R}, indicating good data quality.  The \textit{Gaia} EDR3 photometry is corrected for extinction using the code from Appendix A of \citet{2021A&A...649A...1G}. The extinction is calculated using Equation 1 and Table 1 from \citet{2018A&A...616A..10G}.

\subsection{APOGEE DR17 Data Selection}
\begin{figure}
    \centering
    \includegraphics[width=\columnwidth]{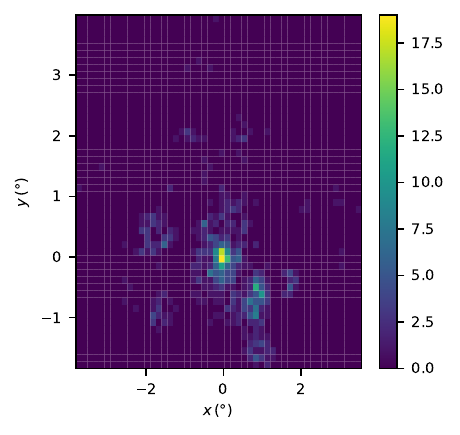}
    \caption{Spatial distribution of sources selected from APOGEE \citep[DR17;][]{2022ApJS..259...35A}.  Pixels are color-coded by the number of the stars within that pixel.}
    \label{fig:apogee_spatial_dist}
\end{figure}

From APOGEE DR17 we select stars within a 4-degree circle centred at $(\alpha,\delta)=(283.76353, -30.477006)$.  We remove stars for which any of the following conditions are met : \verb|STAR_BAD| (bit position $23$ in \verb|ASPCAPFLAG|) is set to $1$, \verb|STARFLAG| is nonzero, or \verb|DUPLICATE| (bit position $4$ in \verb|EXTRATARG|) is set to $1$. The APOGEE DR17 catalogue is cross-matched with \textit{Gaia} EDR3, allowing us to apply to the APOGEE sample the same proper motion selection that we adopt for the \textit{Gaia} EDR3 sample (Sec~\ref{sec:gaia_data_selection}).  Our selections reduce the number of stars in our APOGEE DR17 sample from 6890 to 683.
Fig.~\ref{fig:apogee_spatial_dist} shows the spatial distribution of stars in the selected APOGEE sample.  Since APOGEE DR17 has only partial coverage in this sky area, we cannot directly combine the APOGEE and \textit{Gaia} samples.  Therefore, below, we choose to analyse the line-of-sight velocity data separately from the sky position and proper motion data.  %
(Sec~\ref{sec:model_los}).

\subsection{CMD mask}
The samples selected according to the above criteria will have contamination from Milky Way stars that belong to neither M54 nor Sgr.  In order to minimise this contamination, we construct an additional mask in the colour/magnitude diagram (CMD), combining a
mask derived from the data and a mask derived from isochrone templates.

To construct the CMD mask from the data, we consider two annuli that are centred on M54, the first with inner and outer radii of $0.03, 0.2 \degree$, respectively, and the second
with radii of $2.0, 4.0 \degree$.  Then, from the selected sources within each annulus, we calculate \mbox{2-D} histograms in colour-magnitude space, with $40$ uniform bins from $-0.2$ to $2$ in the $BP-RP$ colour
dimension and $40$ uniform bins from $13$ to $19$ in the $G$ magnitude dimension (only for the purpose of constructing the CMD masks, we relax the magnitude limit of $G< 17.3$ that we apply when analysing the samples).  Both 2-D histograms are divided by the annulus area, so that the result indicates the density of stars.  We then subtract the histogram corresponding to the outer annulus from the histogram corresponding to the inner annulus.  Since the outer annulus should contain a larger fraction of Milky Way contaminants, we expect this difference to be more positive in colour/magnitude bins that are less contaminated by the Milky Way.  We mask sources within colour/magnitude bins for which the difference histogram has value $<20$ stars per degree\textsuperscript{2}. In order to decrease the effect of noise, we perform a `closing' operation, using a  $2 \times 2$ matrix of ones as the structuring element on the CMD mask. 

To construct the CMD mask from isochrone templates, we consider three sets of theoretical isochrones that are motivated by the age/metallicity relations derived from MUSE spectroscopy of M54 by \citet{2019ApJ...886...57A}: young metal-rich, with [Fe/H]=$-0.04$ and age between $1.4-2.8$ Gyr, 2) intermediate-age metal-rich, with [Fe/H]=$-0.29$ and age between $1-7.8$ Gyr, and old metal-poor, with [Fe/H]=-1.41 and age between $9.4-15$ Gyr.  For each set, we obtain the corresponding isochrones in Gaia passbands from the Dartmouth Stellar Evolution Database (DSED) \citep{2008ApJS..178...89D}.  We sample the age ranges with a step size of $0.2$ Gyr.  Shifting each isochrone for a fiducial distance modulus of 17.27 \citep{2007ApJ...667L..57S},  we then mask all CMD bins (using the same binning scheme adopted above for the data-derived mask) that do not contain any of the sampled isochrones.

\begin{figure*}
    \centering
    \includegraphics[width=\textwidth]{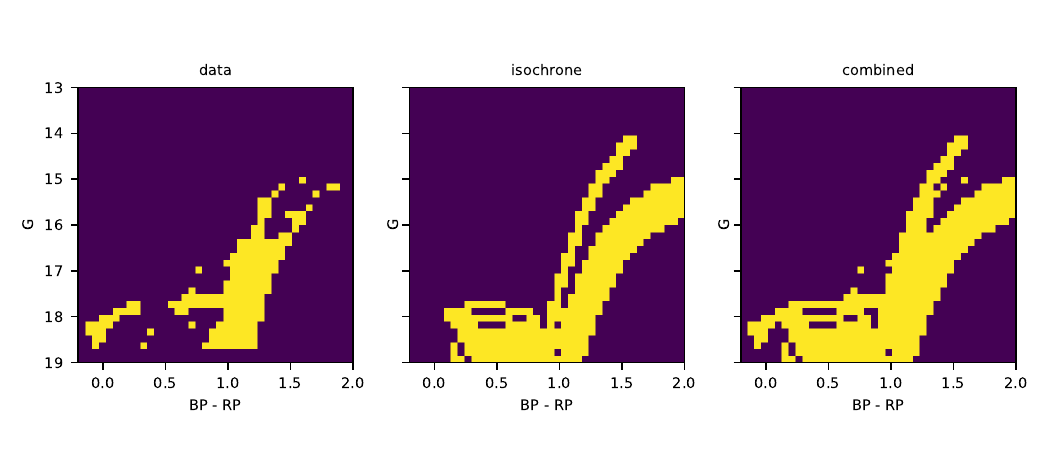}
    \caption{CMD mask for data selection, with yellow/blue indicating unmasked/masked. The left panel is the empirical CMD mask derived directly from the data; the centre panel is the CMD mask derived from theoretical isochrones; the right panel is the combined CMD mask used in our analysis. }
    \label{fig:cmd_mask_combine}
\end{figure*}

Finally, we combine data-derived and isochrone-derived masks by masking only those bins that are masked in both schemes. 
 Fig~\ref{fig:cmd_mask_combine} shows the data-derived, isochrone-derived and combined CMD masks.  Application of the combined mask reduces the number of stars in our selected \textit{Gaia} EDR3 sample from 44337 to 34265.

\section{Mixture Model for the Sgr/M54 system}
\label{sec:method}

We assume the sample of unmasked stars includes contributions from Sgr, M54 and the Milky Way.  First we construct and fit a mixture model for the joint 4D distribution of \textit{Gaia}-observed positions and proper motions.

\subsection{Projected position}
\label{sec:spatial_pm_formula}

For M54 and Sgr, we assume that the 2D positions, 
\begin{equation}
    \matr{s}=\begin{pmatrix} x \\ y \end{pmatrix},
\end{equation}
of stars are drawn independently from analytic surface density profiles that allow for elliptical symmetry about centers specified by
\begin{equation}
    \matr{s_0}=\begin{pmatrix} x_0 \\ y_0 \end{pmatrix}.
\end{equation}
For ellipticity $\epsilon\equiv 1-b/a$ specified by semi-major and semi-minor axes $a$ and $b$, and with semi-major axis pointing along position angle $\theta_0$ (increasing northward of the zero-point at due East), a star's `elliptical radius', $R_e$, relative to the center satisfies 
\begin{equation}
    R_e^2 = \matr{R}^\top \matr{R}=\left( \matr{s}-\matr{s_0}\right)^{\top} \matr{R}_{\theta}^{\top} \begin{pmatrix}
            1 & 0 \\ 0 & \frac{1}{\left(1-\epsilon\right)^2} \\
        \end{pmatrix} \matr{R}_{\theta} \left( \matr{s}-\matr{s_0}\right),
    \label{eq:r_elliptical}
\end{equation}
where  
\begin{equation}
    \matr{R}_{\theta}=\begin{pmatrix} \cos(\theta) & \sin(\theta) \\ -\sin(\theta) & \cos(\theta) \end{pmatrix}\\
\end{equation}
is the rotation matrix.  

We assume that within the observed region, the 2D spatial distribution of stars belonging to M54 is drawn from the probability distribution specified by \citet{1962AJ.....67..471K}, 
\begin{equation}
    f_{s,\text{King}}(\matr{s} \mid \Theta_{\text{King}} )=k_{\text{King}} \left(\frac{1}{\sqrt{1+(R_e/r_c)^2}}-\frac{1}{\sqrt{1+(r_t/r_c)^2}} \right)^2,
\label{eq:spatial_pdf_king}
\end{equation}
with $\Theta_{\text{King}}\equiv\{x_0,y_0,r_c,r_t,\epsilon,\theta \}$ the vector of free parameters that specify the centroid, `core' and `tidal' radii, ellipticity and position angle, respectively.  The normalisation constant $k_{\text{King}}$ guarantees that $\int_{\rm field} f_s(x, y)\,dx\,dy = 1$, where the integration integrates only over the sampled field.

 We assume that the 2D spatial distribution of stars belonging to Sgr is drawn from the probability distribution specified by \citet{1911MNRAS..71..460P} 
\begin{equation}
    f_{s,\text{Plummer}}(\matr{s} \mid \matr{\Theta}_{\text{Plummer}})=\frac{k_{\text{Plummer}}}{(1+R_e^2/r_p^2)^2},
\label{eq:spatial_pdf_plummer}
\end{equation}
with $\Theta_{\text{Plummer}}\equiv\{x_0,y_0,r_p,\epsilon,\theta \}$ the vector of free parameters that specify the centroid, Plummer scale radius, ellipticity and position angle, respectively, and $k_{\text{Plummer}}$ the normalisation constant.

We assume the Milky Way contaminants are drawn from a probability distribution specified by a first order polynomial function, 
\begin{equation}
    f_{s,n}(\matr{s} \mid \Theta_{L})=k_{L} \left|1 +  b_0 x + b_1 y \right|
\label{eq:spatial_pdf_poly}
\end{equation}
with ${\Theta_{L}\equiv(b_0, b_1)}$ the vector of free parameters that specify the polynomial coefficients, and $k_L$ the normalisation constant.

\subsection{Proper motion}
For each of the M54, Sgr and Milky Way components, we assume that the proper motions, 
\begin{equation}
    \matr{\mu}=\begin{pmatrix} \mu_{x} \\ \mu_{y} \end{pmatrix},
\end{equation}
are drawn independently from bivariate normal probability distributions with centers
\begin{equation}
    \matr{\mu_0}(x,y \mid \matr{A})=\begin{pmatrix} \mu_{0_x} \\ \mu_{0_y} \end{pmatrix}+  \matr{A} \begin{pmatrix} x \\ y \end{pmatrix},
    \label{eq:bivariatenormal}
\end{equation}
where 
\begin{equation}
    \matr{A}=\begin{pmatrix} a_1 & a_2 \\ a_3 & a_4 \end{pmatrix}
\end{equation}
specifies dependence on projected sky position, and covariance matrix
\begin{equation}
    \matr{\Sigma}=\matr{R}_{\psi}^{\top} \matr{S} \matr{R}_{\psi},
\end{equation}
where
\begin{equation}
    \matr{S}\equiv\begin{pmatrix} d_{0}^2 & 0 \\ 0 & d_{1}^2 \end{pmatrix}
\end{equation}
and 
\begin{equation}
    \matr{R}_{\psi}\equiv\begin{pmatrix} \cos(\psi) & -\sin(\psi) \\ \sin(\psi) & \cos(\psi) \end{pmatrix},
\end{equation}
with $\psi$ corresponding to the position angle of the long axis of the proper motion ellipsoid (increasing from the positive direction of proper motion in right ascension direction toward  the negative direction of proper motion in declination direction).

For a given component, then, the observed proper motion has probability density
\begin{equation}
    f_{\mu}(\matr{\mu} \mid \Theta_{\mu})=\frac{\mathcal{N}(\matr{\mu} \mid \matr{\mu_0},\matr{\Sigma}+\matr{\Sigma_0})}{C},
    \label{eq:fmu}  
\end{equation}
where $\matr{\Theta_{\mu}}\equiv \{\mu_{0_x},\mu_{0_y},a_1,a_2,a_3,a_4,d_0,d_1,\psi \}$ specifies free parameters and $\matr{\Sigma_0}$ is the covariance matrix of the observed proper motion coordinates  (adopted from the Gaia EDR3 catalog), and 
\begin{equation}
    C\equiv \displaystyle\int_{\mu_{x_1}}^{\mu_{x_2}}\displaystyle\int_{\mu_{y_1}}^{\mu_{y_2}} \mathcal{N}(\matr{\mu} \mid \matr{\mu_0},\matr{\Sigma}+\matr{\Sigma_0})\,d\mu_x\,d\mu_y
\end{equation}
is a normalizing factor that accounts for the finite proper motion selection window (Section \ref{sec:gaia_data_selection}), chosen to ensure $\iint f_{\mu}(\matr{\mu} \mid \matr{\Theta}_{\mu})\, d\matr{\mu}=1$.  

\subsection{Line-of-sight Velocity}
\label{sec:model_los}

We assume that the line-of-sight velocities, $v$, sampled by APOGEE are drawn independently from a mixture that includes contributions from Sgr, M54 and the Milky Way, all of which are assumed to have velocities that follow univariate normal probability distributions.  That is, for each of the three components we assume the APOGEE-observed velocities are drawn from 
\begin{equation}
    f_{v}(v \mid  \matr{\Theta}_{v}) = \mathcal{N}(v \mid v_0 + e_{x} x + e_{y} y,\sigma^2+\delta_v^2),
    \label{eq:rv_gaussian}
\end{equation}
where the free parameters in $\matr{\Theta}_{v}$ specify a mean velocity, $v_0+e_{x}x+e_{y}y$, that can vary linearly with sky position and has central value $v_0$, and intrinsic velocity dispersion $\sigma$.   $\delta_v$ is the observational error associated with the velocity measurement (adopted from the APOGEE catalogue).  In practice, we assume $e_{x}=e_{y}=0$ for the M54 and MW components, allowing for a non-zero line-of-sight velocity gradient only in Sgr.

\subsection{Likelihood Functions}
\label{sec:likelihood}
With the normalized probability distributions specified for observed position and proper motion coordinates, the 4D observation of position and proper motion coordinates has joint probability density
\begin{equation}
    \begin{split}
        f_{\matr{s},\matr{\mu}}(\matr{s},\matr{\mu} \mid \Theta)= \sum_{i=1}^{N_{\rm comp}} \phi_i\,f_{s}(\matr{s}\mid\matr{\Theta}_{\matr{s}_i}) f_{\mu}(\matr{\mu}\mid\matr{\Theta}_{\matr{\mu}_i}),
    \end{split}
    \label{eq:mixture_model_spatial_pm_pdf}
\end{equation}
where $\matr{\Theta}$ specifies all free parameters and the sum is over the $N_{\rm comp}$ different mixture components, each of which has contribution weighted by $\phi_i$, the fraction of the sample that is contributed by the $i^{th}$ component.  For each of the $N_{\rm comp}=3$ components, we specify $\phi_i$ in terms of parameters $p_i$, with the latter defined according to

\begin{equation}
    \phi_i = \begin{cases}
        \begin{aligned}
            p_i \left(1-\sum_{j=1}^{i-1} \phi_j \right) &= p_i\prod_{j=1}^{i-1}(1-p_j) & i < N_{\rm comp} \\
            1-\sum_{j=1}^{i-1} \phi_j &= \prod_{j=1}^{i-1}(1-p_j) & i=N_{\rm comp}
        \end{aligned}
    \end{cases}
    \label{eq:pmember}
\end{equation}

As described in Section \ref{sec:spatial_pm_formula}, for different components we adopt different analytic functions for the probability density of stellar positions.  For M54 we use $f_{\matr{s}}(\matr{s}\mid\matr{\Theta}_{\matr{s}})=f_{\matr{s},\text{King}}(\matr{s}\mid\matr{\Theta}_{\matr{\text{King}}})$; for Sgr we use $f_{\matr{s}}(\matr{s}\mid\matr{\Theta}_{\matr{s}})=f_{\matr{s},\text{Plummer}}(\matr{s}\mid\matr{\Theta}_{\matr{\text{Plummer}}})$; for the Milky Way we use $f_{\matr{s}}(\matr{s}\mid\matr{\Theta}_{\matr{s}})=f_{\matr{s,n}}(\matr{s}\mid\matr{\Theta}_{\matr{L}})$.  

For all three components, we assume that the observed proper motions follow the general form of $f_{\matr{\mu}}(\matr{\mu}\mid\matr{\Theta}_{\matr{\mu}})$ given by Equation \ref{eq:fmu}.  For both M54 and the Milky Way, we assume that the proper motion centre, $\matr{\mu_0}$, is independent of the spatial position within the sampled region, with $a_1=a_2=a_3=a_4=0$.  For Sgr, following \citet{2020MNRAS.497.4162V}, we allow for a linear dependence of the mean proper motion on projected sky position, letting the components of $\matr{A}$ be nonzero.

Assuming the observations of different sources are uncorrelated, the 4D Gaia sample of $N=34265$ stars has likelihood
\begin{equation}
    \mathcal{L}_4=\prod_{k=1}^{N} f(\matr{s}_k,\matr{\mu}_k \mid \matr{\Theta})
\end{equation}

Similarly, the 1D Apogee sample of $N_v=683$ stars has likelihood
\begin{equation}
    \mathcal{L}_{1} = \prod_{k=1}^{N_v} \sum_{i=1}^{N_{\rm comp}} q_{k,i}f_{v}(v_k \mid  \matr{\Theta}_{v, \rm i}),
    \label{eq:qki}
\end{equation}
where the weight
\begin{equation}
    q_{k,i} = \frac{c_{i} f_{\matr{s},\matr{\mu}}(\matr{s}_k,\matr{\mu}_k\mid \matr{\Theta}_i) }{\sum_{j=1}^{N_{\rm comp}} c_{j} f_{\matr{s},\matr{\mu}}(\matr{s}_k,\matr{\mu}_k\mid\matr{\Theta}_j)}
    \label{eq:rv_single_star}
\end{equation}
is the probability of the $k^{\rm th}$ star's membership in the $i^{\rm th}$ component, given the \textit{Gaia}-observed position and proper motion parameters $\matr{\Theta}_i$ of the 4D model for the $i^{\rm th}$ component.  The three parameters $c_1$, $c_2$ and $c_3$ represent the mixing fractions within the spectroscopic sample. 
 We use the analogue of Equation~\ref{eq:pmember} to specify the $c_i$ in terms of two independent free parameters $m_1, m_2$, requiring that $\sum_{i=1}^{N_{\rm comp}}c_i=1$.

\subsection{Inference}
From the 4D and 1D data sets, denoted $D_4$ and $D_1$, respectively, we are interested in inferring posterior probability distributions
\begin{equation}
    f_4(\matr{\Theta} \mid D_4)=\frac{\mathcal{L}_4\,\pi(\matr{\Theta})}{\int {\mathcal{L}_4\,\pi(\matr{\Theta})\,d\matr{\Theta}}}
\end{equation} 
and 
\begin{equation}
    f_1(\matr{\Theta}_v \mid D_1)=\frac{\mathcal{L}_1\,\pi(\matr{\Theta}_v)}{\int \mathcal{L}_1\,\pi(\matr{\Theta}_v)\,d\matr{\Theta}_v},
\end{equation}
where $\pi(\matr{\Theta})$ and $\pi(\matr{\Theta}_v)$ are priors and the denominator on the right-hand side of each equation is the marginalized likelihood, or `evidence'.  In order to estimate $f_1(\matr{\Theta}_v\mid D_1)$, when evaluating the component membership probabilities in Equation~\ref{eq:qki} we hold 4D model parameters $\matr{\Theta}$ fixed at the maximum a posteriori (MAP) value obtained when estimating $f_4(\matr{\Theta}\mid D_4)$. 
 Table~\ref{tab:prior_posterior_mean_std_new} lists the priors that we adopt for all model parameters.  For all parameters except M54's core radius and those that specify member fractions, we adopt uniform priors over finite ranges.  For the core radius of M54, since our sample excludes the core region (Figure \ref{fig:spatial2d_center}), we adopt a truncated Gaussian prior that is centred on the previous measurement by \citet{1996AJ....112.1487H} (2010 edition) which is $0.0015\degree$. The reported measurement does not include uncertainty, so we use $0.0015\degree$  as the standard deviation of the Gaussian prior to allow a broad distribution. We also set a lower bound $r_c>1\times 10^{-4}$ degree to ensure numerical stability of the calculation of the normalisation factor in the King profile.  For the parameters $p_i$ (and $m_i$) that specify member fractions in the 4D (1D) model according to Equation \ref{eq:pmember}, we adopt a Dirichlet prior over ($\phi_1, \ldots, \phi_{N_{\rm comp}}$) and ($c_1,\ldots,c_{ N_{\rm comp}}$) which can be interpreted as the uniform distribution under the constraints $\sum_i \phi_i = 1, \phi_i>0$ and $c_i = 1, c_i>0$.  

Independently for each data set, we obtain random samples from the posteriors using the software package emcee \citep{emcee}, which implements a Markov chain Monte Carlo (MCMC) method.  When fitting the 4D sample, we use 3000 walkers, 5000 steps and thinning parameter of 2.  When fitting the 1D sample, we use 500 walkers, 5000 steps and thinning parameter of 1.
 In both cases, we discard the first half of the chain as burn-in.

\begin{table*}
    \centering
    \caption{Free parameters of the model fit to 4D Gaia sample of projected position and proper motion, including priors, quantiles at $(0.1587, 0.5, 0.8413)$ and maximum a posteriori (MAP).}
    \label{tab:prior_posterior_mean_std_new}
    \begin{tabular}{ccccc}
        & Prior & Params & $(0.1587, 0.5, 0.8413)$ Quantile & MAP \\
        \multirow{2}{*}{Member Fraction}&\multirow{2}{*}{$\mathrm{Dirichlet}(1,1,1)$\footnotemark}&$p_{1}$ & $(0.01555,0.01696,0.01837)$        & $0.0179$ \\
        &&$p_{2}$ & $(0.970,0.972,0.974)$ & $0.9724$ \\
        \hline
        \multirow{6}{*}{M54 Spatial ($\matr{\Theta}_{\matr{s},\mathrm{M54}}$)}& $\mathrm{U}[-0.03, 0.03]$ &$x_0\, (\degree)$ & $(-0.003968,-0.001318,0.001358)$ & $0$ \\
        & $\mathrm{U}[-0.03, 0.03]$ &$y_0\, (\degree)$ & $(-0.004626,-0.001924,0.0007959)$ & $-0.0030$ \\
        & $\mathcal{N}(0.0015, 0.0015^2)$ &$r_c\, (\degree)$ & $(0.0007815,0.001938,0.003343)$ & $0.0017$ \\
        & $\mathrm{U}[0.03, 10.0]$ &$r_t\, (\degree)$ & $(6.483,8.274,9.499)$ & $9.7$ \\
        & $\mathrm{U}[-0.5\pi, 0.5\pi]$ &$\theta\, (\text{rad})$ & $(-0.8814,-0.1925,0.686)$ & $-0.24$ \\
        & $\mathrm{U}[0.1, 1.0]$ &$1-\epsilon$ & $(0.8612,0.9326,0.9803)$ & $0.88$ \\
        \hline
        \multirow{5}{*}{M54 proper motion ($\matr{\Theta}_{\matr{\mu},\mathrm{M54}}$)}& $\mathrm{U}[-4.32, -1.32]$ &${\mu_0}_{x}\, (\text{mas yr}^{-1})$ & $(-2.689,-2.682,-2.676)$ & $-2.6779$ \\
        & $\mathrm{U}[-3.01, -0.02]$ &${\mu_0}_{y}\, (\text{mas yr}^{-1})$ & $(-1.386,-1.380,-1.374)$ & $-1.3769$ \\
        & $\mathrm{U}[0.0, 2.0]$ &${s_0}\, (\text{mas yr}^{-1})$ & $(0.04562,0.05254,0.05951)$ & $0.0542$ \\
        & $\mathrm{U}[0.0, 2.0]$ &${s_1}\, (\text{mas yr}^{-1})$ & $(0.03546,0.04439,0.05283)$ & $0.0411$ \\
        & $\mathrm{U}[0.0, 0.5\pi]$ &${\psi}\, (\text{rad})$ & $(0.3421,0.8044,1.263)$ & $0.87$ \\
        \hline
        \multirow{2}{*}{MW Spatial ($\matr{\Theta}_{\matr{s},\mathrm{MW}}$)}& $\mathrm{U}[-0.3, 0.3]$ &$b_{0}\, (\text{deg}^{-1})$ & $(-0.1619,-0.1318,-0.1018)$ & $-0.128$ \\
        & $\mathrm{U}[-0.3, 0.3]$ &$b_{1}\, (\text{deg}^{-1})$ & $(0.05190,0.08108,0.1100)$ & $0.101$ \\
        \hline
        \multirow{5}{*}{MW proper motion ($\matr{\Theta}_{\matr{\mu},\mathrm{MW}}$)}& $\mathrm{U}[-9.0, 5.0]$ &${\mu_0}_{x}\, (\text{mas yr}^{-1})$ & $(-7.276,-3.646,1.393)$ & $-2.1$ \\
        & $\mathrm{U}[-20.0, 0.0]$ &${\mu_0}_{y}\, (\text{mas yr}^{-1})$ & $(-16.78,-12.34,-7.502)$ & $-9.6$ \\
        & $\mathrm{U}[1.0, 17.0]$ &${s_0}\, (\text{mas yr}^{-1})$ & $(3.841,7.398,13.59)$ & $15.0$ \\
        & $\mathrm{U}[1.0, 17.0]$ &${s_1}\, (\text{mas yr}^{-1})$ & $(2.609,3.745,5.937)$ & $2.7$ \\
        & $\mathrm{U}[0.0, 0.5\pi]$ &${\psi}\, (\text{rad})$ & $(0.1999,0.515,1.189)$ & $0.41$ \\
        \hline
        \multirow{5}{*}{Sgr Spatial ($\matr{\Theta}_{\matr{s},\mathrm{Sgr}}$)}& $\mathrm{U}[1.0, 20.0]$ &$r_h\, (\degree)$ & $(5.013,5.093,5.174)$ &  $5.083$ \\
        & $\mathrm{U}[-3.0, 3.0]$ &$x_0\, (\degree)$ & $(0.2580,0.2856,0.3136)$ & $0.252$ \\
        & $\mathrm{U}[-3.0, 3.0]$ &$y_0\, (\degree)$ & $(-0.08488,-0.0733,-0.06153)$ & $-0.063$ \\
        & $\mathrm{U}[-0.5\pi, 0.5\pi]$ &$\theta\, (\text{rad})$ & $(-0.2785,-0.2725,-0.2665)$ & $-0.2704$ \\
        & $\mathrm{U}[0.1, 1.0]$ &$1-\epsilon$ & $(0.3928,0.3987,0.4046)$ & $0.4008$ \\
        \hline
        \multirow{9}{*}{Sgr proper motion ($\matr{\Theta}_{\matr{\mu},\mathrm{Sgr}}$)}& $\mathrm{U}[-1.0, 1.0]$ &$a_1\, (\mathrm{mas\ yr^{-1} \ deg^{-1}})$ & $(0.007206,0.007693,0.008193)$ & $0.00782$ \\
        & $\mathrm{U}[-1.0, 1.0]$ &$a_2\, (\mathrm{mas\ yr^{-1} \ deg^{-1}})$ & $(0.004597,0.005317,0.006028)$ &  $0.00515$ \\
        & $\mathrm{U}[-1.0, 1.0]$ &$a_3\, (\mathrm{mas\ yr^{-1} \ deg^{-1}})$ & $(-0.02349,-0.02305,-0.02262)$ &  $-0.02305$ \\
        & $\mathrm{U}[-1.0, 1.0]$ &$a_4\, (\mathrm{mas\ yr^{-1} \ deg^{-1}})$ & $(-0.01427,-0.01364,-0.01299)$ & $-0.01315$ \\
        & $\mathrm{U}[-4.32, -1.32]$ &${\mu_0}_{x}\, (\text{mas yr}^{-1})$ & $(-2.680,-2.679,-2.678)$ & $-2.67913$ \\
        & $\mathrm{U}[-3.01, -0.02]$ &${\mu_0}_{y}\, (\text{mas yr}^{-1})$ & $(-1.395,-1.394,-1.393)$ & $-1.39374$ \\
        & $\mathrm{U}[0.0, 2.0]$ &${s_0}\, (\text{mas yr}^{-1})$ & $(0.1037,0.1044,0.1051)$ & $0.10433$ \\
        & $\mathrm{U}[0.0, 2.0]$ &${s_1}\, (\text{mas yr}^{-1})$ & $(0.1483,0.1492,0.1502)$ & $0.14856$ \\
        & $\mathrm{U}[0.0, 0.5\pi]$ &${\psi}\, (\text{rad})$ & $(0.9237,0.9351,0.9466)$ & $0.938$ \\
        \hline
    \end{tabular}
\end{table*}

\section{Results}
\label{sec:results}

Random samples sampled from posterior distributions are publicly available at \url{https://doi.org/10.5281/zenodo.10659516}.  Table~\ref{tab:prior_posterior_mean_std_new} and Table~\ref{tab:prior_posterior_mean_std_vlos} summarise the marginalised 1D posterior for each model parameter in the 4D and 1D mixture models, respectively, identifying the quantile at $(0.1587, 0.5, 0.8413)$ for each parameter. The member fraction ($\phi_i$) for each component in the 4D mixture model, taking the posterior mean, is $95.55\%$ for Sgr, $2.75\%$ for MW\@, and $1.70\%$ for M54. The member fraction ($c_i$) for each component in the 1D mixture model, taking the posterior mean, is $71.65\%$ for Sgr, $2.38\%$ for MW\@, and $25.97\%$ for M54. We note that the tidal radius $r_t$ for M54 tends to be very large in the posterior distribution. We have tried to enforce a strong Gaussian prior using the literature value $0.165\degree$ \citep[2010 edition]{1996AJ....112.1487H} as the mean, which will restrict the range of posterior distribution around $0.165\degree$, but we do not observe any improvement in the residual distribution or any change in our results. If we take the MAP estimation, the probability mass of normalized King profile within radial distance $1\degree \leq r \leq 8\degree$ is $13\%$. The distribution is still dominated by the central area. With these considerations, we decide to use the less informative prior which will lead to a large tidal radius. 

We note first that our estimate of the posterior PDF for the Milky Way component of the proper motion distribution is not well converged.  This is due to the relatively small number of MW stars that survive our initial proper motion selection (Figure \ref{fig:pm_selection}; the inferred contribution of MW stars to the 4D model is $\lesssim 3\%$).

\begin{figure*}
    \centering        
    \includegraphics[width=\textwidth]{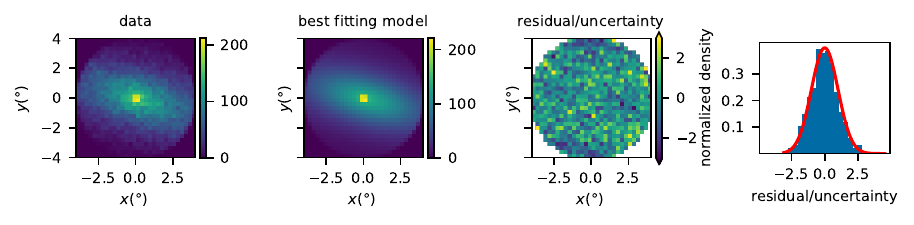}
    \caption{Projected density of stars in the selected sample (left), best-fitting model (middle left), residuals normalised by Poisson uncertainties (middle right), and normalized distribution of the residuals/uncertainty. The colour bars indicate the number of stars in each bin (left, middle left) and residual/uncertainty in the corresponding bin (middle right right). The red curve in the right plot is the normal distribution centred at $0$ with variance $1$.}
    \label{fig:2d_surface_density}
\end{figure*}

\begin{figure}
    \centering
    \includegraphics[width=\columnwidth]{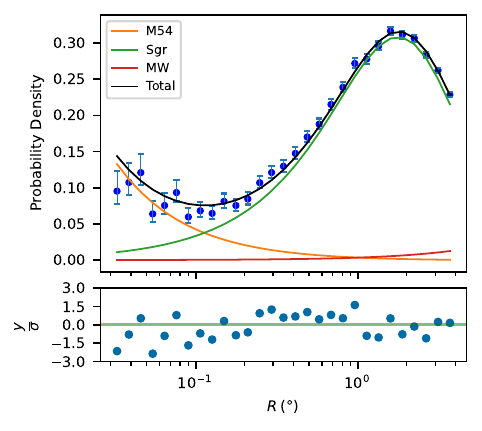}
    \caption{\textit{Top:} Probability density of circular radial coordinate $R_{\rm circ}=\sqrt{x^2+y^2}$.   
    Points with errorbars are the data.  Overplotted is our best-fitting model, with 
    individual contributions from M54, Sgr and Milky Way components indicated by colored curves.  \textit{Bottom:} residuals normalized by Poisson uncertainties.}
    \label{fig:1d_surface_density}
\end{figure}

Figure \ref{fig:2d_surface_density} displays the distribution of projected positions in the data (left panel), best-fitting model (middle-left panel), residuals normalised by Poisson uncertainties estimated from the expected values under the best-fitting model (middle-right panel), and the histogram of the normalised residuals (right panel).  Figure \ref{fig:1d_surface_density} displays the probability distribution for circular radial coordinate $R_{\rm circ}=\sqrt{x^2+y^2}$, with individual contributions from the modelled M54, Sgr and MW components indicated.  Here the observed data are binned for illustrative purposes only, as the models are fit to the discrete data for stellar position. 
 The lower sub-panel in Figure \ref{fig:1d_surface_density} shows residuals, normalised by the Poisson uncertainty. 
 We find that our mixture model provides a reasonably good fit, with normalised residuals approximately following a unit-normal distribution (right panel of Figure \ref{fig:2d_surface_density} and bottom panel of Figure \ref{fig:1d_surface_density}).  %

\begin{figure*}
    \centering
    \includegraphics[width=\textwidth]{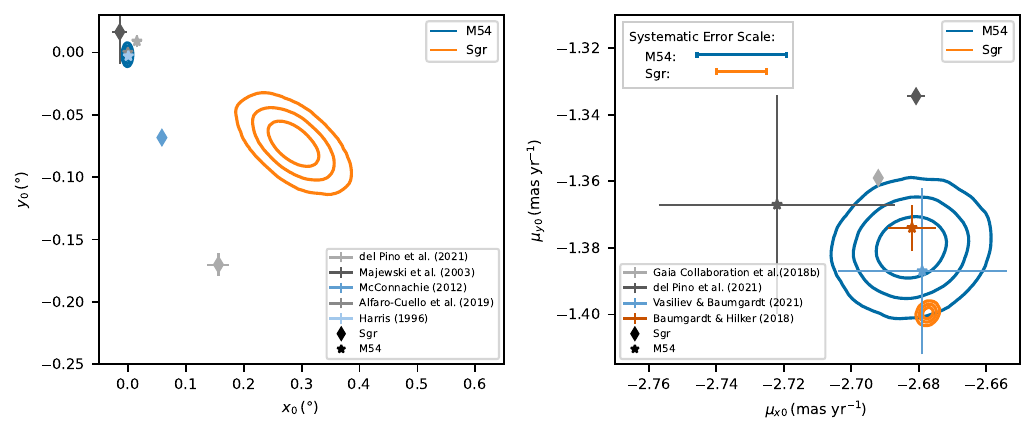}
    \caption{Contours enclosing 68\%, 95\% and 99\% of posterior probability for model parameters specifying projected centroids (left) and mean proper motion coordinates (right) of M54 and Sgr components.  For the Sgr component, for which our proper motion model has mean varying linearly with position, the plotted mean is calculated at the centre of Sgr.  Markers indicate previously-published measurements (filled circles for M54, filled stars for Sgr). The top left lines in the right plot indicate the expected magnitude of systematic error in the mean proper motion estimated for each object \citep{2021MNRAS.505.5978V}.}
    \label{fig:spatial_pm_center_contour}
\end{figure*}

The contours in Figure~\ref{fig:spatial_pm_center_contour} represent posterior PDFs that we estimate for spatial centroids (left panel) of M54 and Sgr, and mean proper motions (right panel); for Sgr, for which we allow the mean proper motion to vary linearly with projected positions from the spatial centroid, the mean proper motion is evaluated at the fitted centroid.
The inferred spatial centres of M54 and Sgr are offset by an angle of $\left| \Delta \matr{s} \right|=0.295 \pm 0.029^{\circ}$---corresponding to a physical separation of $0.135 \pm 0.013$ kpc at the adopted distance of $26.28$ kpc---with M54 projected to the northwest of the Sgr centre.  

The inferred mean (central) proper motions of M54 and Sgr are offset by $\Delta\matr{\mu}= [0.0049, -0.0197] \pm [0.0068, 0.0062] $ mas yr$^{-1}$, which are $[0.61, -2.46] \pm [0.85, 0.77] $ km s$^{-1}$ at the adopted distance of $26.28$ kpc, with M54 moving more slowly than Sgr toward the south. %
However, the precision of mean proper motions estimated for extended objects is limited by spatial covariance within the \textit{Gaia}'s EDR3 proper motion catalog \citep{2021A&A...649A...2L}.  We use Equation 2 from \citet{2021MNRAS.505.5978V} to estimate the magnitude of this systematic error `floor', which depends on the angular size of the object in question.  Using the core radius $r_c = 0.0017\degree$ for M54 and half-light radius $r_h=5.083\degree$ for Sgr to represent the relevant angular scales of these objects, the corresponding systematic errors are $\epsilon_{\text{sys}}\approx 0.026$ mas yr$^{-1}$ for M54 and $\epsilon_{\text{sys}}\approx 0.015$ mas yr$^{-1}$ for Sgr.  Thus the offset that we infer between the mean proper motions of M54 and Sgr is similar to the expected contribution from systematic error. 
Thus the offset that we infer between the mean proper motions of M54 and Sgr has a similar magnitude to the expected systematic error.

\begin{figure}
    \centering
    \includegraphics[width=\columnwidth]{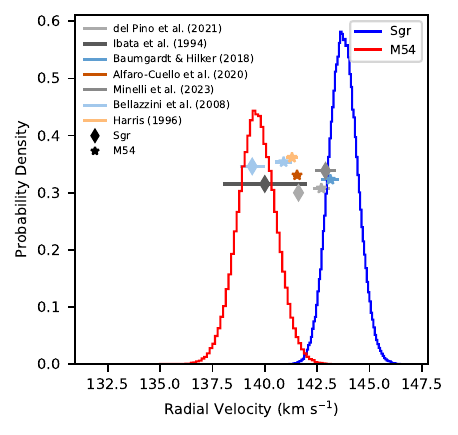}
    \caption{Posterior probability distribution of the mean line-of-sight velocity of Sgr (blue) and M54 (red), from our mixture modeling of our sample selected from the Apogee (DR17) survey.  Dots and stars represent previously-published measurements for Sgr and M54,  respectively, with error bars representing $1\sigma$ uncertainty.}
    \label{fig:radial_velocity_core_m54_hist}
\end{figure}

Fig~\ref{fig:radial_velocity_core_m54_hist} displays posterior PDFs for mean line-of-sight velocities of M54 and Sgr.  We infer mean values of $v_0=139.63 \pm 0.92 \,\mathrm{km~s^{-1}}$ for M54 and $v_0=143.74 \pm 0.69 \,\mathrm{km~s^{-1}}$ for Sgr (evaluated at the inferred centroid of Sgr), for an offset of $\Delta v_0=4.1\pm 1.2$ km s$^{-1}$.  We infer a significant gradient in Sgr's line-of-sight velocities, $e_x=-1.68\pm 0.57$ km s$^{-1}$ deg$^{-1}$ and $e_y=-4.80 \pm 0.74 $ km s$^{-1}$ deg$^{-1}$, implying a gradient of magnitude $\sim 5$ km s$^{-1}$ deg$^{-1}$ ($\sim 10$ km s$^{-1}$ kpc$^{-1}$), similar to the gradient previously reported by \citet{2021ApJ...908..244D}.

Taken at face value, these results suggest that M54 and Sgr are offset in phase space, with the detected offsets significant at the $10\sigma$ and $3.4\sigma$ levels in projected sky position and line-of-sight velocity components, respectively.  The inferred proper motion offset, while significant at the $\sim 3\sigma$ level based on our posterior PDF alone, is similar to the expected level of systematic error.

\begin{table*}
    \centering
    \caption{Free parameters of model fit to 1D APOGEE sample of line-of-sight velocity, including priors and quantiles at $(0.1587, 0.5, 0.8413)$.}
    \label{tab:prior_posterior_mean_std_vlos}
    \begin{tabular}{cccc}
        & Prior &  Params & $(0.1587, 0.5, 0.8413)$ Quantile \\
        \multirow{2}{*}{Member Fraction} & \multirow{2}{*}{$\mathrm{Dirichlet}(1,1,1)$} & $m_1$ & $(0.6047,0.7206,0.8242)$ \\
        & & $m_2$ & $(0.03337,0.06656,0.1336)$ \\
        \hline
        \multirow{4}{*}{Sgr $v_{\rm LOS}$} & $\mathrm{U}[100, 200]$ & $v_0 \, (\mathrm{km\ s^{-1}})$ & $(143.2,143.9,144.6)$ \\
        & $\mathrm{U}[10, 100]$ & $\sigma \, (\mathrm{km\ s^{-1}})$ & $(12.48,12.91,13.37)$ \\
        & $\mathrm{U}[-10, 10]$ & $e_x \, (\mathrm{km\ s^{-1}\ deg^{-1}}) $ & $(-2.252,-1.675,-1.115)$ \\
        & $\mathrm{U}[-10, 10]$ & $e_y \, (\mathrm{km\ s^{-1}\ deg^{-1}}) $ & $(-5.534,-4.798,-4.064)$ \\
        \hline
        \multirow{2}{*}{MW $V_{\rm LOS}$} & $\mathrm{U}[-50, 50]$ & $v_0 \, (\mathrm{km\ s^{-1}})$ & $(-30.07,-11.62,8.933)$ \\
        & $\mathrm{U}[10, 150]$ & $\sigma \, (\mathrm{km\ s^{-1}})$ & $(49.34,62.69,80.41)$ \\
        \hline
        \multirow{2}{*}{M54 $V_{\rm LOS}$} & $\mathrm{U}[100, 200]$ & $v_0 \, (\mathrm{km\ s^{-1}})$ & $(138.7,139.6,140.5)$ \\
        & $\mathrm{U}[1, 100]$ & $\sigma \, (\mathrm{km\ s^{-1}})$ & $(8.153,8.907,9.630)$ \\
    \end{tabular}
\end{table*}

\section{Discussion}
\label{sec:discussion}

\subsection{Comparison to Previous work}

For comparison with our inferences of the centroids and mean motions of M54 and Sgr, Figures \ref{fig:spatial_pm_center_contour} and \ref{fig:radial_velocity_core_m54_hist} indicate previous measurements found in the literature.  Our inference for the spatial centre of M54 is consistent with those reported by \citet{2021ApJ...908..244D, 2019ApJ...886...57A, 1996AJ....112.1487H}. However, our inference for the spatial centre of Sgr is offset, by $\sim 15$ arcmin, from the center listed in the review by \citet[][which uses the value originally reported by \citealt{1994Natur.370..194I}]{2012AJ....144....4M}, and by similar amounts from measurements reported by \citet{2003ApJ...599.1082M} and \citet{2021ApJ...908..244D}.  Notably, \textit{none} of the listed measurements of Sgr's centre agree with each other.  This situation is not entirely surprising, given the different instrumental sensitivities and complex mix of stellar populations near Sgr's centre.  The measurement by \citet{2003ApJ...599.1082M} coincides with a steep `cusp' that they detect at the position of M54, which might reflect contamination of the M giant sample by the relatively young and metal-rich M54 component reported in later work \citep{2019ApJ...886...57A}.  Indeed \citet{2021ApJ...908..244D} find that the inclusion of M54 stars in their sample would shift the fitted centroid of Sgr toward M54.  Their measurement that we show in Figure \ref{fig:spatial_pm_center_contour} uses a sample that masks a circular region of radius $0.16^{\circ}$ centred on M54, chosen to excise the cluster within its King tidal radius (for comparison, our central mask described in Section \ref{sec:gaia_data_selection} has radius $0.03^{\circ}$).  However, \citet{2021ApJ...908..244D} downplay the resulting offset from M54, noting the potential for systematic errors due to the \textit{Gaia} scanning pattern that is apparent in their chosen sample from Gaia DR2.  Thus ours is the first inference of an offset between the spatial centres of M54 and Sgr that is based on a mixture model that simultaneously accounts for both objects within the same data set.  Given our selection of a relatively bright magnitude limit ($G<17.3$; Section \ref{sec:gaia_data_selection}), our data does not show an visible scanning pattern, and thus our result of offset should be less susceptible to systematic errors arising from the spatial dependence of \textit{Gaia's} faint magnitude limit.

Our inference for M54's proper motion vector is consistent with other recent \textit{Gaia}-based measurements reported by \citet{2018MNRAS.478.1520B,2021ApJ...908..244D, 2021MNRAS.505.5978V}.  However, our inference for Sgr's proper motion (evaluated at Sgr's centre) is statistically inconsistent with those reported by \citet{2018A&A...616A..12G, 2021ApJ...908..244D}, with our measurement indicating a larger component toward the south.  Again we see that all reported measurements for Sgr disagree with each other.  \citet{2018A&A...616A..12G} note that their PM measurements are affected by \textit{Gaia}'s scanning pattern and varying astrometric incompleteness in \textit{Gaia} DR2, which contributes to a systematic error that they estimate to be $\sigma_{sys} \sim 0.030 \, \text{mas yr}^{-1}$ and $\sim 0.036 \, \text{mas yr}^{-1}$ in $\mu_{\alpha} \cos \delta$ and $\mu_{\delta}$, respectively, similar to the offset between their result and ours.  The measurements by \citet{2018A&A...616A..12G} and \citet{2021ApJ...908..244D} are  based on the \textit{Gaia} DR2 catalogue and the proper motion zero-points are both different and spatially varying between the DR2 and DR3 catalogue and this  systematic error may account for the offset compared to our result.  In any case, even taking any of the available measurements for Sgr's proper motion vector at face value, one finds a similarly significant offset (albeit with reversed direction in the case of the previous measurements based on \textit{Gaia} DR2) with respect to our inference for the proper motion of M54.  Again, however, we note that the apparent offset is similar in magnitude to the expected contribution from systematic error.%

Finally, Fig~\ref{fig:radial_velocity_core_m54_hist} shows that previously published measurements of the mean line-of-sight velocities of M54 and Sgr are scattered between the values we infer.  For M54, our estimate (evaluated at the inferred centre of Sgr) is in reasonable agreement with those reported by \citet{1996AJ....112.1487H}, \citet{2008AJ....136.1147B} and \citet{2020ApJ...892...20A}, but more discrepant (with our measurement having a smaller line-of-sight velocity) with those of \citet{2018MNRAS.478.1520B} and \citet{2021ApJ...908..244D}.  For Sgr, our result agrees well with that of \citet{2023A&A...669A..54M}, marginally with those reported by \citet{2012AJ....144....4M} and \citet{2021ApJ...908..244D}, and poorly (with our measurement being $\sim 4$ km s$^{-1}$ more positive) with that of \citet{2008AJ....136.1147B}.  This scatter likely reflects not only systematic errors due to the different zero-points of the different instrumental setups, but also the different criteria used to select and/or identify M54 and Sgr samples.  For example, \citet{2023A&A...669A..54M} select their Sgr sample from fields observed outside the central region containing M54, with a broad CMD filter chosen to pass a wide range of age and metallicity.  Other studies \citep[e.g.]{2008AJ....136.1147B} define M54 and/or Sgr samples by applying relatively blue and red filters along the red giant branch sequences; as previously discussed (Section \ref{sec:intro}), this selection can confuse a young, metal-rich component in M54 \citep{2019ApJ...886...57A} with Sgr, perhaps helping to explain why many of the previous results lie between the mean velocities we measure for the two objects.  Again we emphasise that our mixture model's separation between M54 and Sgr components is based only on the observed distribution of phase-space coordinates, and is agnostic regarding the colour/magnitude and hence age/metallicity properties of each object.

\subsection{Robustness to sample selection and modelling assumptions}

The most significant offset we detect between Sgr and M54 is between their 2D spatial centroids.  We now examine the extent to which this offset may be driven by our choice of modelling assumptions and sample selection.  As described in Section \ref{sec:gaia_data_selection}, we selected our sample of stellar positions (and proper motions) to be spread over a circular field of radius $4^{\circ}$ (excluding the crowded region within a field of radius $0.03^{\circ}$) centred on M54.  Our model assumes that within this region, the stellar surface density of Sgr is well described by an elliptical Plummer profile.  However, as a result of ongoing tidal interaction with the Milky Way, Sgr is known to display distorted morphology in its outer regions, ultimately extending to tidal tails that encircle the sky \citep{mateo96,2003ApJ...599.1082M}.  To date, the most detailed view of Sgr's internal stellar structure and kinematics comes from the \textit{Gaia}-based study by \citet{2021ApJ...908..244D}, who report a central bar-like feature extending out to the tidal tails, and an overall triaxial morphology, with a slowly rotating ($V_{\rm rot}\sim 4$ km s$^{-1}$) central region of radius $\sim 500$ pc that is embedded within an outer envelope that appears to be expanding along its longest axis.  

\begin{figure}
    \centering
    \includegraphics[width=\columnwidth]{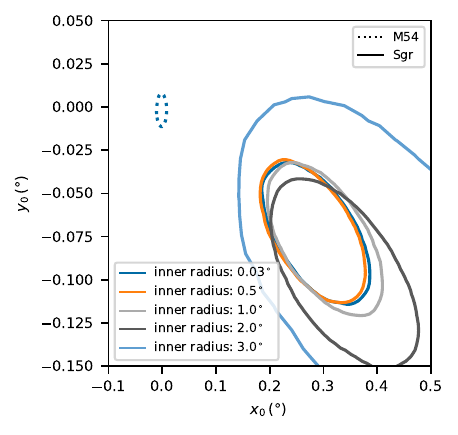}
    \caption{$3\sigma$ level contour of the posterior probability distributions for the spatial centres of M54 (dotted) and Sgr (solid).  Contours are color-coded by the inner radius of the sky coverage for different adopted samples.}
    \label{fig:posterior_contour_different_inner_radii}
\end{figure}

In order to check whether our inference for Sgr's spatial centre is sensitive to the omitted data in the centre of the M54, Fig~\ref{fig:posterior_contour_different_inner_radii} shows $3\sigma$ contours (i.e., enclosing 99.7\% of the probability mass) from posterior PDFs for the centroids of M54 and Sgr, obtained using samples excluding fields of radius $0.03^{\circ}$ (the original choice), $0.5^{\circ}$, $1^{\circ}$, $2^{\circ}$ and $3^{\circ}$. When the inner radius is larger than $0.03^{\circ}$, we remove the M54 component entirely from the mixture model because we expect M54's contribution to the sample to be negligible. The contours change only slightly until the inner radius reaches $\ga 2^{\circ}$; contours from cases with larger inner radius enclose the contours from the cases with smaller inner radius. This result indicates that the spatial centre we infer for Sgr is robust to our choice of the size of the removal area at the centre of M54.

\begin{figure}
    \centering
    \includegraphics[width=\columnwidth]{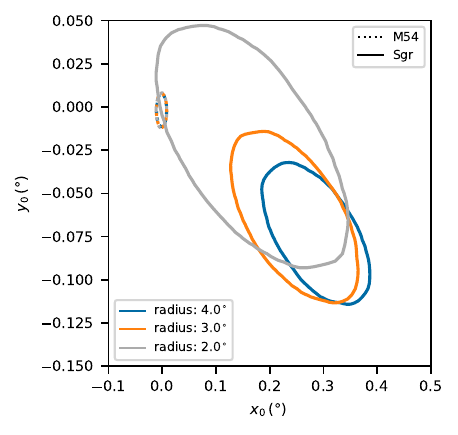}
    \caption{$3\sigma$ level contour of the posterior probability distributions for the spatial centres of M54 (dotted) and Sgr (solid).  Contours are color-coded by the radius of the sky coverage for different adopted samples.}
    \label{fig:posterior_contour_different_radii}
\end{figure}

In order to gauge the effect of Sgr's complicated morphology on our inference about its centre, we repeat our fits using samples selected over smaller fields (still centred on M54).  Fig~\ref{fig:posterior_contour_different_radii} shows $3\sigma$ contours from posterior PDFs for the centroids of M54 and Sgr, obtained using samples spanning fields of radius $4^{\circ}$ (the original choice), $3^{\circ}$ and $2^{\circ}$.  Unsurprisingly, all cases return statistically identical inferences for the centroid of M54.  For Sgr, contours for all cases continue to enclose the centroid inferred from our original sample; however, as the field size decreases, the contours expand and we see a systematic reduction in the inferred offset from M54.  For a field radius of $2^{\circ}$, the offset is detected only marginally, with the  $3\sigma$ contour overlapping the inferred centroid of M54.  

The inflation of statistical uncertainty as the field radius decreases is unsurprising, as the sample becomes more confined to the region well within Sgr's fitted Plummer radius ($R_p\sim 5^{\circ}$), where the 2D density is close to uniform.  The systematic shift in the inferred Sgr centroid, toward M54 as the field radius decreases, is similar in magnitude to the increase in statistical error.  Thus the systematic component cannot be neglected, and we must acknowledge that the significance with which we detect a spatial offset between M54 and Sgr is vulnerable to model mismatch, particularly with regard to the stellar density distribution toward the outer regions of Sgr.  However, residuals with respect to best-fitting models (Figures \ref{fig:2d_surface_density} and \ref{fig:1d_surface_density}) do not reveal obvious signs of model mismatch.  

In order to examine more quantitatively the potential for asymmetric morphology within Sgr to induce spurious detection of an offset from M54, we fit an alternative model for Sgr's 2D spatial configuration that explicitly includes an asymmetric component.

The alternative asymmetric model is constructed by modifying Equation~\ref{eq:r_elliptical}.  The original matrix $\matr{R}$ from Equation~\ref{eq:r_elliptical} is

\begin{equation}
    \matr{R}=\begin{pmatrix}
            1 & 0 \\ 0 & \frac{1}{\left(1-\epsilon\right)} \\
        \end{pmatrix} \matr{R}_{\theta} \left( \matr{s}-\matr{s_0}\right)
    \label{eq:r_elliptical_original}
\end{equation}
Now we construct the asymmetric model to stretch the distribution along just one of the semi-major axes:

\begin{equation}
    \matr{R'}=\begin{pmatrix}
        \frac{\lvert\matr{R}_{00}\rvert (1-s_p)}{2 s_{p}} + \frac{\matr{R}_{00} (1+s_p)}{2 s_{p}}  \\ \matr{R}_{10} \\
        \end{pmatrix}
    \label{eq:r_elliptical_asym}
\end{equation}

\noindent where $\matr{R}_{ij}$ represents the element of the $i$th row and $j$th column from $\matr{R}$ (index starts from $0$). This expression ensures that

\begin{equation}
    \matr{R'}_{00} = \begin{cases}
        \matr{R}_{00} &\quad \text{if } \matr{R}_{00}>=0 \\
        \matr{R}_{00}/s_p &\quad \text{if } \matr{R}_{00}<0. \\
    \end{cases}
    \label{eq:r_elliptical_asym_explain}
\end{equation}
We construct the asymmetric model by replacing the $R_e^2 = \matr{R}^\top\matr{R}$ in Equation~\ref{eq:spatial_pdf_plummer} with $R_e^2 = \matr{R'}^\top\matr{R'}$.  The normalisation factor $k_P$ in Equation~\ref{eq:spatial_pdf_plummer} is recalculated using the new expression. The new parameter $s_p$ controls how much one side of the Sgr spatial distribution is elongated. The prior on the parameter $s_p$ is $\mathrm{U}[1, 5]$.

We fit the asymmetric model using the software package Dynesty \citep{2020MNRAS.493.3132S, sergey_koposov_2023_8408702, 2004AIPC..735..395S, 2019S&C....29..891H, 2009MNRAS.398.1601F}, which returns an estimate of the marginalised likelihood, or `evidence', as well as a random sample from the posterior.  The evidence provides a metric for model selection.  

\begin{figure}
    \centering
    \includegraphics[width=\columnwidth]{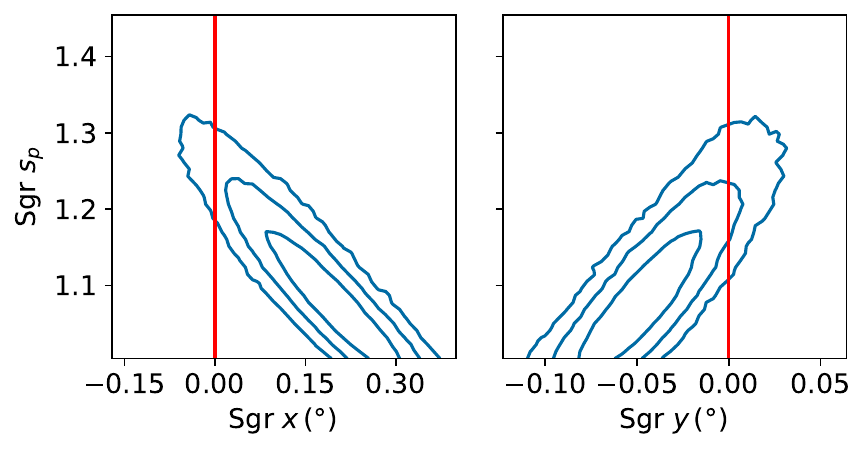}
    \caption{Contours enclosing 68\%, 95\% and 99\% of posterior probability for model parameters specifying the one-sided `stretch' factor of an asymmetric model for Sgr, and the $x$, $y$ components of Sgr's centroid.  Red vertical lines mark the origin, which is chosen to coincide with the center of M54 (from \citealt{2019ApJ...886...57A}.}
    \label{fig:asym_scale_center_hist2d}
\end{figure}

The contours in Figure~\ref{fig:asym_scale_center_hist2d} represent posterior PDFs for the asymmetric stretch factor, $s_p$, and the inferred centroid of Sgr.  The observed degeneracy indicates that for sufficiently large $s_p$, near the $2\sigma$ contour of our posterior PDF, Sgr's centroid can shift to coincide with the center of M54 at $(x,y)=(0,0)$.  However, comparing results from our asymmetric and original symmetric model, the natural logarithm of the evidence ratio is $3.13\pm 0.45$, in favor of the simpler, symmetric model.  In order to interpret this numerical result, we perform a simulation in which we fit our symmetric and asymmetric models to mock data generated under our best-fitting (to the real data) symmetric model.  In this case we know the input model has no inherent asymmetry, and we obtain a similar evidence ratio, which has a logarithm of $3.03\pm 0.44$, again in favor of the symmetric model.  We conclude that, while the offset we infer between centroids of M54 and Sgr can potentially be caused by asymmetry in Sgr's morphology, the real data do not give reason to prefer at least our simple asymmetric model over the original symmetric one.

\subsection{Colour/magnitude distribution}
\begin{figure*}
    \centering
    \includegraphics[width=\textwidth]{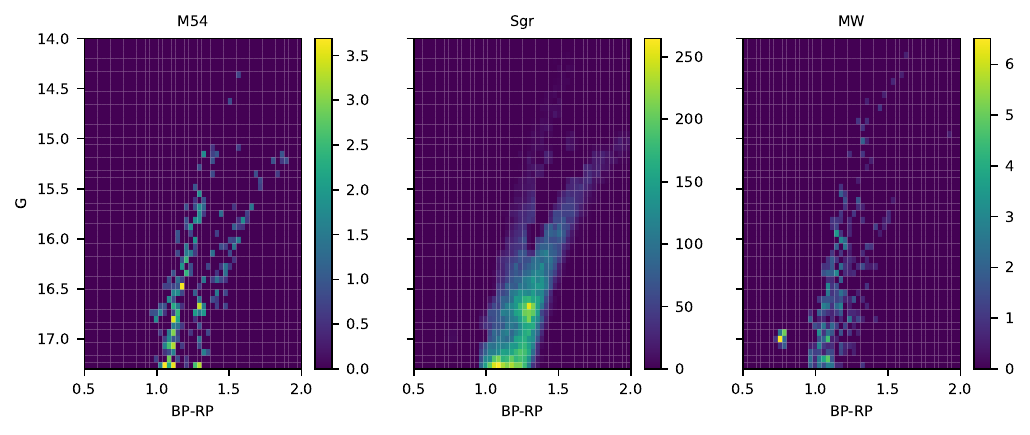}
    \caption{Color/magnitude distributions for stars inferred to have $>50\%$ probability of membership in the M54 (left), Sgr (middle) and MW (right) components.  The colour bar indicates the number of stars in each bin.}
    \label{fig:cmd_per_member}
\end{figure*}

We reiterate that our mixture model is agnostic regarding the colour/magnitude distribution of stars belonging to the individual components of M54, Sgr and the Milky Way.  This represents a departure from most previous analyses of M54 and/or Sgr, which typically analyse separate samples that are pre-selected from relatively blue and red sequences defined along the red giant branch \citep[e.g.,][]{2003ApJ...599.1082M,2008AJ....136.1147B}.  In contrast, our inferences of mixing fraction and structural/kinematic parameters for each mixture component are informed only by the observed stellar positions and motions.  Afterwards, however, we can use the inferred parameters and mixing fractions to examine the corresponding colour/magnitude distribution of each component.  For each of M54, Sgr and the Milky Way, Figure \ref{fig:cmd_per_member} displays the number of stars per pixel in colour/magnitude space having probability of membership $> 50\%$ in that component.  

We see that the Milky Way contribution tends to be fainter and bluer than the M54 and Sgr populations, only sparsely populating the red giant sequences.  Sgr stars lie primarily to the red side of our CMD filter, with a prominent red clump visible at $G\sim 16.7$.
Finally, we see that while M54 is composed primarily of stars along the blue ridge of our CMD filter, it has a population of high-probability members along the redder sequence as well.  This result is broadly consistent with that of \citet{2019ApJ...886...57A}, who use MUSE spectroscopy to separate the M54 region into an old (12.2 Gyr), metal-poor ([Fe/H]$=-1.41$) population a young (2.2 Gyr), metal-rich ([Fe/H]$=-0.04$) population and an intermediate-age (4.3 Gyr), metal-rich ([Fe/H]$=-0.29$) population, with the first two being more centrally concentrated than the last.  It is this central concentration that associates the young, metal-rich population with M54 in our model. 
  \citet{2019ApJ...886...57A} speculate that these stars may have formed in situ in the nuclear region, in response either to the funnelling of gas during Sgr's most recent pericentric passage of the Milky Way, or to the sinking of enriched gas to the centre of a sufficiently massive M54.  

\subsection{Orbits of M54 and Sgr}
If M54 is offset from the centre of Sgr in phase space, then it becomes interesting to trace the history of the M54/Sgr interaction, which will depend on the masses and internal structure of both objects as well as the external tidal field.  In order to explore at least some aspects of this dependence, we consider here a simplified `toy' model in which M54 is a point mass and Sgr is embedded within an extended dark matter halo that follows the flexible `CoreNFW' (cNFW) density profile of \citet{2018MNRAS.481..860R}.  This spherically-symmetric halo model is defined by the enclosed-mass profile
\begin{equation}
    M_{\rm cNFW}(r)=f^nM_{\rm NFW}(r),
\end{equation}
where
\begin{equation}
    M_{\rm NFW}(r)=M_{200}\frac{\ln(1+r/r_s)-r/r_s(1+r/r_s)^{-1}}{\ln(1+c_{200})-c_{200}(1+c_{200})^{-1}}
\end{equation}
is the `NFW' profile that characterizes halos produced in cosmological N-body simulations that assume `cold' dark matter \citep{1996ApJ...462..563N,1997ApJ...490..493N}.  Here, $M_{200}=M_{\rm NFW}(r_{200})$ is a proxy for halo mass, with $r_{200}$ the radius of the sphere inside which the mean density is $3M_{200}/(4\pi r^3_{200})=200\rho_c$, where $\rho_c=3H_0^2/(8\pi G)$ is the critical density of the Universe (we assume the Hubble constant has value $H_0=70$ km s$^{-1}$ Mpc$^{-1}$).  With $r_{200}$, the scale radius $r_s$ defines halo concentration, $c_{200}\equiv r_{200}/r_s$.
The function
\begin{equation}
    f^n\equiv \biggl [\tanh\biggl (\frac{r}{r_c}\biggr )\biggr ]^n
\end{equation}
allows the `cuspy' ($d\log\rho/d\log r=-1$) inner behavior of the NFW profile to be transformed, to a degree controlled by parameter $0<n\leq 1$, to a `core' of uniform density within a sphere of radius $r_c$.

In order to examine sensitivity to the structure of Sgr's dark matter halo, we consider two possibilities, both of which assume that M54 is a point particle of mass $M_{\rm M54}=1.41\times 10^{6}\,\mathrm{M_{\odot}}$ \citep{2018MNRAS.478.1520B}  and Sgr consists of a stellar component, described by a spherical Plummer model, embedded within a cNFW dark matter halo.
In the first case, the cNFW halo has $r_c=0$, which simplifies to the original NFW profile.  In the second case, the cNFW halo has a fully-formed ($n=1$) core of radius, $r_c=r_{p}$, that is assumed to equal the Plummer radius fit to the observed stellar density profile \citep{2016MNRAS.459.2573R}.
In both cases, we adopt a dark matter halo concentration of $c_{200}=10$, and for Sgr a total stellar mass (within the bound Plummer sphere) of $M_{*,\rm Sgr}=2\times 10^{7} \,\mathrm{M_{\odot}}$ \citep[][we find no sensitivity to different plausible choices for Sgr's stellar mass, such as the value of $M_{*,\rm Sgr}=1\times 10^{8} \,\mathrm{M_{\odot}}$ from \citealt{2020MNRAS.497.4162V}]{2012AJ....144....4M}.
For both dark matter profiles, we choose $M_{200}$ so that the dark matter mass enclosed within the Plummer radius matches dynamical mass estimated as $M(r_{\text{half}}) = 2.5 r_{\text{p}} \sigma_{v}^2/G$ \citep{2009ApJ...704.1274W}, where we adopt $r_p=5.092^{\circ}$ as the MAP result from our modeling and $\sigma_{v}=11.4$ km s$^{-1}$ is the velocity dispersion \citet{1997AJ....113..634I}.  For the NFW and cNFW profiles, this requirement gives $M_{200}=1.29 \times 10^9 \,\mathrm{M_{\odot}}$ and $M_{200}=2.10 \times 10^9 \,\mathrm{M_{\odot}}$, respectively.  
In order to account for any effect of the Large Magellanic Cloud (LMC) on the orbits of M54 and Sgr, we include an LMC potential that we model using a Hernquist profile with total mass $1.38 \times 10^{11}\;
    \mathrm{M_{\odot}}$ and scale radius $16.09 \; \mathrm{kpc}$
\citep{2019MNRAS.487.2685E}; however, we find that the inclusion or exclusion of the LMC potential has negligible effect on our results.  

We sample the present-day 2D position, proper motions and line-of-sight velocities for both M54 and Sgr from the posterior PDFs inferred from our mixture models.
We include the systematic errors \citep{2021MNRAS.505.5978V} in the proper motion samples by adding Gaussian noise to the proper motion components sampled from our posterior PDF, accounting for uncertainty `floors' of $\epsilon_{\text{sys}}= 0.026\ \text{mas yr}^{-1}$ for the proper motion components of M54 and $\epsilon_{\text{sys}}=0.015\ \text{mas yr}^{-1}$ for Sgr.
We draw present-day line-of-sight distances to M54 and Sgr independently from a normal distribution with mean 26.28 kpc and standard deviation 0.33 kpc, corresponding to the measurement by \citet{2021MNRAS.505.5957B} of the distance to M54.  %
For Sgr, assumed here to be spherically symmetric, we sample the Plummer radius from our posterior PDF for $r_h$.  For the LMC, we adopt the present-day phase space coordinates from \citet{2018A&A...616A..12G}.%

Given the stated initial conditions, we use the software package \texttt{gala} \citep{adrian_price_whelan_2022_7299506} to calculate the orbits of M54, Sgr and the LMC within a Milky Way potential that adopts \texttt{gala}'s MilkyWayPotential class, which assumes a disk model from \citet{2015ApJS..216...29B}.  We choose \texttt{gala}'s RK5Integrator, which uses a fifth order Runge-Kutta method.  In order to gauge the effects of dynamical friction on M54 from both Sgr and the MW and on Sgr and the LMC from the MW, we calculate orbits with and without using galpy's ChandrasekharDynamicalFrictionForce class \citep{2015ApJS..216...29B}, which implements the Chandrasekhar dynamical friction force following \citet{2016MNRAS.463..858P}.  For each combination of NFW vs. cNFW for Sgr's dark matter profile and with/without dynamical friction implemented, we draw $100$ samples from our posterior PDFs that provide initial conditions, and then calculate the corresponding orbits over the past 1 Gyr. The timestep is $5\,\text{Myr}$ and the number of steps is $200$.  

\begin{figure*}
    \centering
    \includegraphics[width=\textwidth]{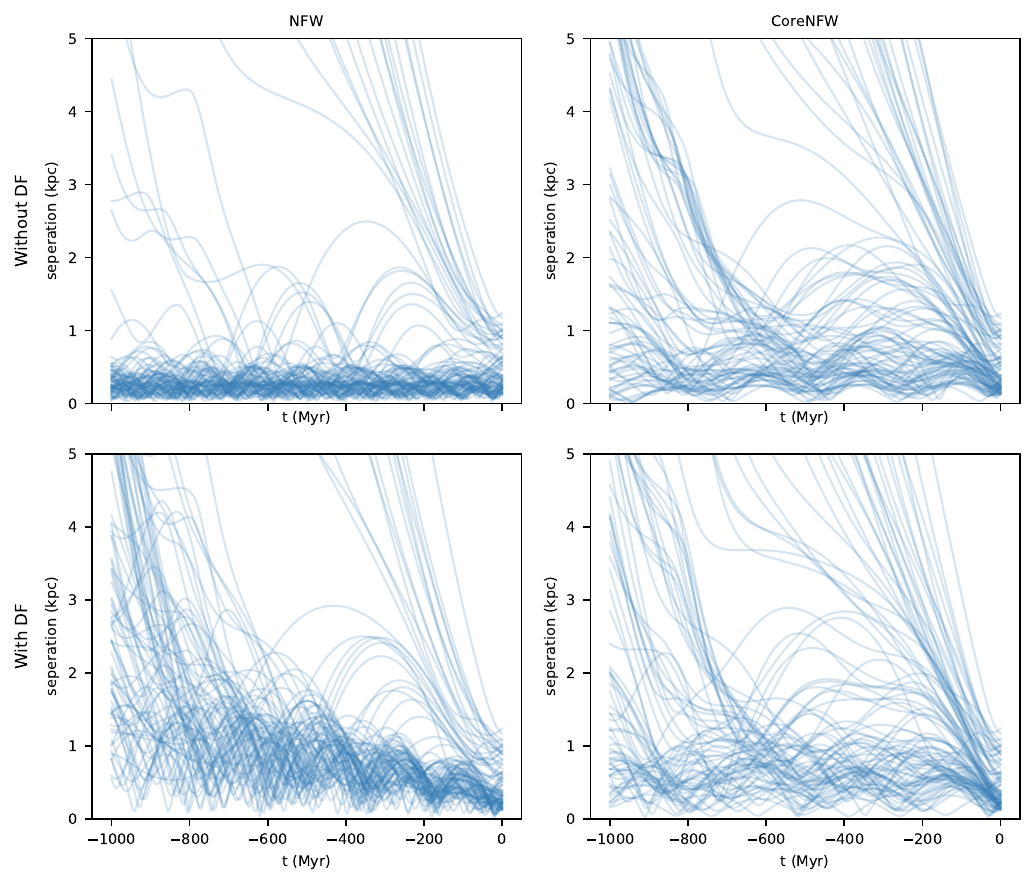}
    \caption{Evolution of the 3D spatial offset between the centers of M54 and Sgr, with orbits integrated backward in time from the present day ($t=0$), with boundary conditions drawn from posterior PDFs of our 5D phase-space models (present-day distances to M54 and Sgr are drawn independently from a normal distribution with mean 26.28 kpc and standard deviation 0.33 kpc, based on the distance measurement of \citet{2021MNRAS.505.5957B}).  Each line represents the orbits calculated from one set of initial conditions drawn from the posterior. 
    Results shown in left/right panels are calculated using NFW/CoreNFW potentials for Sgr's dark matter halo.  Orbits shown in the top/bottom rows neglect/include an implementation of dynamical friction.  }
    \label{fig:orbit_seperation}
\end{figure*}

For each computed orbit, Fig~\ref{fig:orbit_seperation} depicts the recent history of the (3D) spatial offset between M54 and the centre of Sgr, tracing back 1 Gyr from the present day.  Left and right columns show results using the NFW and coreNFW model, respectively, for Sgr's dark matter halo; top and bottom panels show results obtained without and with dynamical friction implemented.

For cases where the present-day offset is $\lesssim 0.7$ kpc, we observe a dependence on the dark matter density profile within Sgr.  If Sgr's dark matter halo follows the NFW profile, we find that most orbits are approaching apocentre, with the M54/Sgr separation increasing from an offset as small as $\sim 0.1$ kpc at the most recent pericentric passage, which occurred between $\sim 15 - 40$ Myr ago.  In the absence of dynamical friction, the orbit remains approximately unchanged for the past Gyr for most of the samples (upper left panel of Figure \ref{fig:orbit_seperation}).  With dynamical friction implemented (lower left panel of \ref{fig:orbit_seperation}), the offset at apocentre has decreased steadily from a value of $\sim 1.5$ kpc over the past 700 Myr. 

In contrast, if Sgr's dark matter halo follows the adopted coreNFW profile, then present-day offsets that are $\lesssim 0.7$ kpc are close to pericentre for most of the samples, after a most recent apocentre $\gtrsim 100$ Myr ago, at which the offset was $\gtrsim 0.5$ kpc (right panels of Figure \ref{fig:orbit_seperation}).  As expected, the implementation of dynamical friction has relatively little effect when the dark matter halo is cored \citep{2006MNRAS.373.1451R, 2009MNRAS.397..709I}, allowing M54 to `wobble' about the centre indefinitely \citep{2012MNRAS.426..601C}.

If the present-day offset is $\gtrsim 0.7$ kpc, and thus dominated by a large offset in line-of-sight distance between M54 and Sgr, all sampled orbits would imply that M54 fell into Sgr within the past 200 Myr, regardless of whether Sgr's dark matter halo follows a NFW or cNFW profile.  However, this scenario seems unlikely, as one expects Sgr to be losing mass, not accreting, during this timeframe that includes Sgr's own pericentric passage about the Milky Way \citep{2021MNRAS.501.2279V}.

We emphasise that our orbit calculations neglect realistic details of the Sgr/M54 system---most notably, Sgr's tidal disruption by the Milky Way.  The orbits that we calculate imply that Sgr's most recent pericentric passage about the Milky Way brought it within $\sim 15$ kpc of the Galactic centre, just $\sim 45$ Myr ago, consistent with the Sgr orbit inferred by \citep{2021MNRAS.501.2279V}.  However, even if we restrict our calculations only to the previous 50 Myr, the qualitative conclusion remains that an M54/Sgr system that has internal offset $\lesssim 0.7$ kpc tends to be appropaching apocentre if Sgr is embedded within an NFW halo, and near pericentre if Sgr is embedded within a cored halo.  A more thorough investigation of the history of the M54/Sgr interaction would require updating previous N-body approaches \citep[e.g.,][]{2008AJ....136.1147B,2023MNRAS.523.2721H} in light of the present-day phase-space coordinates presented above.%

\section{Summary \& Conclusions}
\label{sec:conclusions}

Motivated by longstanding questions regarding the formation and orbital histories of nuclear star clusters, here we have analyzed the 5D (sky positions and proper motions from \textit{Gaia} EDR3, line-of-sight velocities from APOGEE DR17) structure and kinematics of the M54/Sgr system, the only example of a NSC/host system for which such multi-dimensional data are available.  In order to obtain simultaneous estimates of the centers of M54 and Sgr in each of these coordinates, we have constructed and fit a mixture model that accounts also for contamination from the Milky Way foreground/background.  Our approach differs from most previous efforts, which consider M54 and/or Sgr samples separately after filtering according to colour/magnitude and/or 2D spatial criteria.  In analysing both M54 and Sgr populations simultaneously within the same sample, our inferences about offsets between the two objects incur fewer systematic errors that arise when comparing estimates made using either independent data sets or independent subsamples of those data sets.

We have reported statistically significant offsets in the spatial centres and line-of-sight velocities of M54 and Sgr. The offset between the spatial centres is $0.295\pm 0.029$ degrees, corresponding to a projected (2D) offset of $0.135\pm 0.013$ kpc at the distance of $26.28$ kpc.  The offset between the line-of-sight velocities is $4.1\pm 1.2$ km s$^{-1}$.  The offset between the PM centres of the M54 and Sgr dSph is $[\Delta\mu_{\alpha}\cos\delta,\Delta\mu_{\delta}]=[0.0049, -0.0197] \pm [0.0068, 0.0062]$ mas~yr$^{-1}$ ($[0.61, -2.46] \pm [0.85, 0.77] $ km s$^{-1}$), comparable to the expected systematic error.  %

By far the most significant offset we infer is between the spatial centres of M54 and Sgr.  However, this result comes with the caveat that the degree of statistical significance is sensitive to our choice of sample field size.  This sensitivity points to the possibility that the inferred offset may be driven by large-scale asymmetry of the spatial distribution of stars within Sgr.  While our fiducial model allows flattened elliptical morphology, we find that the inferred offset between M54 and Sgr can be reduced if we further allow for bilateral asymmetry.  However, we find that standard criteria for model selection disfavour our asymmetric model over the fiducial one.  We conclude, then, that the present data are reasonably well fit by our fiducial model, with M54 significantly offset from Sgr, and do not favour an alternative in which M54 is located exactly at the centre of an asymmetric Sgr host.

Based on our estimates of the 5D phase-space coordinates of M54 and Sgr at the present day, we also investigated the recent orbital history of the interacting pair.  Treating M54 as a point mass and neglecting mass lost from Sgr due to tidal disruption, we calculate orbits within a potential that includes both objects as well as the Milky Way and LMC\@.  We find that if Sgr is embedded in a cuspy (NFW) dark matter halo, then the M54/Sgr orbit is currently most likely approaching apocentric when the present-day offset is $\lesssim 0.7$ kpc, having decayed under dynamical friction from earlier apocentres $\gtrsim 1$ kpc.
In contrast, if Sgr's dark matter halo has a central core, then the M54/Sgr orbit is near pericentre if the present-day offset is $\lesssim 0.7$ kpc.
If the present-day offset between M54 and Sgr is $\gtrsim 0.7$ kpc, and thus dominated by an offset in the line-of-sight distance, the calculated orbits would imply that M54 fell into Sgr within the past 200 Myr regardless of whether Sgr's dark matter halo is cored or cusped.

Finally, the results of our mixture model imply that in addition to a dominant population of old, metal-poor stars, the M54 component includes a population of young, metal-rich stars.  This finding is consistent with the MUSE spectroscopy of \citet{2019ApJ...886...57A}, who associate a young, metal-rich and spatially compact stellar population with M54, and supports the suggestion by \citet{2019ApJ...886...57A} that these stars may have formed in response to the sinking of enriched gas into the center of a massive M54, perhaps triggered during Sgr's most recent pericentric passage of the Milky Way.  Further exploration of this scenario in future work will require N-body simulations, which can be constrained by the results presented above.

\section*{Acknowledgements}
This work is supported by National Science Foundation (NSF) grants AST-1909584 and AST-2206046.  This work made use of the Q3C software \citep{Koposov2006}, NumPy \citep{harris2020array}, SciPy \citep{2020SciPy-NMeth}, emcee \citep{emcee}, Numba \citep{10.1145/2833157.2833162}, gala \citep{gala, adrian_price_whelan_2022_7299506}, Dynesty \citep{2020MNRAS.493.3132S, sergey_koposov_2023_8408702, 2004AIPC..735..395S, 2019S&C....29..891H, 2009MNRAS.398.1601F}, galpy \footnote{http://github.com/jobovy/galpy} \citep{2015ApJS..216...29B} and Astropy\footnote{http://www.astropy.org} which is a community-developed core Python package and an ecosystem of tools and resources for astronomy \citep{astropy:2013, astropy:2018, astropy:2022}.

This work has made use of data from the European Space Agency (ESA) mission
{\it Gaia} (\url{https://www.cosmos.esa.int/gaia}), processed by the {\it Gaia}
Data Processing and Analysis Consortium (DPAC,
\url{https://www.cosmos.esa.int/web/gaia/dpac/consortium}). Funding for the DPAC
has been provided by national institutions, in particular the institutions
participating in the {\it Gaia} Multilateral Agreement.

Funding for the Sloan Digital Sky 
Survey IV has been provided by the 
Alfred P. Sloan Foundation, the U.S. 
Department of Energy Office of 
Science, and the Participating 
Institutions. 

SDSS-IV acknowledges support and 
resources from the Center for High 
Performance Computing  at the 
University of Utah. The SDSS 
website is www.sdss4.org.

SDSS-IV is managed by the 
Astrophysical Research Consortium 
for the Participating Institutions 
of the SDSS Collaboration including 
the Brazilian Participation Group, 
the Carnegie Institution for Science, 
Carnegie Mellon University, Center for 
Astrophysics | Harvard \& 
Smithsonian, the Chilean Participation 
Group, the French Participation Group, 
Instituto de Astrof\'isica de 
Canarias, The Johns Hopkins 
University, Kavli Institute for the 
Physics and Mathematics of the 
Universe (IPMU) / University of 
Tokyo, the Korean Participation Group, 
Lawrence Berkeley National Laboratory, 
Leibniz Institut f\"ur Astrophysik 
Potsdam (AIP),  Max-Planck-Institut 
f\"ur Astronomie (MPIA Heidelberg), 
Max-Planck-Institut f\"ur 
Astrophysik (MPA Garching), 
Max-Planck-Institut f\"ur 
Extraterrestrische Physik (MPE), 
National Astronomical Observatories of 
China, New Mexico State University, 
New York University, University of 
Notre Dame, Observat\'ario 
Nacional / MCTI, The Ohio State 
University, Pennsylvania State 
University, Shanghai 
Astronomical Observatory, United 
Kingdom Participation Group, 
Universidad Nacional Aut\'onoma 
de M\'exico, University of Arizona, 
University of Colorado Boulder, 
University of Oxford, University of 
Portsmouth, University of Utah, 
University of Virginia, University 
of Washington, University of 
Wisconsin, Vanderbilt University, 
and Yale University.

\section*{Data Availability}
The data underlying this article are available in \url{https://doi.org/10.5281/zenodo.10659516}. The datasets were derived from the \textit{Gaia} EDR3 and the APOGEE DR17. The the \textit{Gaia} EDR3 is publicly available at \url{https://www.cosmos.esa.int/web/gaia/earlydr3}, and the APOGEE DR17 is publicly available at \url{https://www.sdss4.org/dr17/}.

\bibliographystyle{mnras}
\bibliography{example} %

\appendix

\bsp	%
\label{lastpage}
\end{document}